\documentclass[10pt,twocolumn,letterpaper]{article}

\usepackage[pagenumbers]{wacv} % To force page numbers, e.g. for an arXiv version

\usepackage{graphicx}
\usepackage{amsmath}
\usepackage{amssymb}
\usepackage{booktabs}

\usepackage{bm}
\usepackage{xcolor,pifont}
\usepackage{multirow}
\usepackage{algorithm}
\usepackage{algpseudocode}
\usepackage{relsize}
\usepackage[accsupp]{axessibility}  % Improves PDF readability for those with disabilities.
\usepackage{color, colortbl}

\newcommand*\colourcheck[1]{%
  \expandafter\newcommand\csname #1mark\endcsname{\textcolor{#1}{\ding{51}}}%
}
\colourcheck{green}

\newcommand*\colourmark[1]{%
  \expandafter\newcommand\csname #1mark\endcsname{\textcolor{#1}{\ding{55}}}%
}
\colourmark{red}

\definecolor{LightCyan}{rgb}{0.88,1,1}

\usepackage[pagebackref,breaklinks,colorlinks]{hyperref}

\usepackage[capitalize]{cleveref}
\crefname{section}{Sec.}{Secs.}
\Crefname{section}{Section}{Sections}
\Crefname{table}{Table}{Tables}
\crefname{table}{Tab.}{Tabs.}

\def\wacvPaperID{1213} % *** Enter the WACV Paper ID here
\def\confName{WACV}
\def\confYear{2024}

\begin{document}

\title{Diffusion in the Dark: \\
A Diffusion Model for Low-Light Text Recognition}

\author{Cindy M. Nguyen\\
Stanford University\\
{\tt\small cindyn@stanford.edu}
\and 
Eric R. Chan \\
Stanford University\\
{\tt\small erchan@stanford.edu}
\and 
Alexander W. Bergman \\
Stanford University\\
{\tt\small awb@stanford.edu}
\and
Gordon Wetzstein\\
Stanford University\\
{\tt\small gordonwz@stanford.edu}
}

\twocolumn[{
\maketitle
\begin{center}
    \captionsetup{type=figure}
    \includegraphics[scale=0.6]{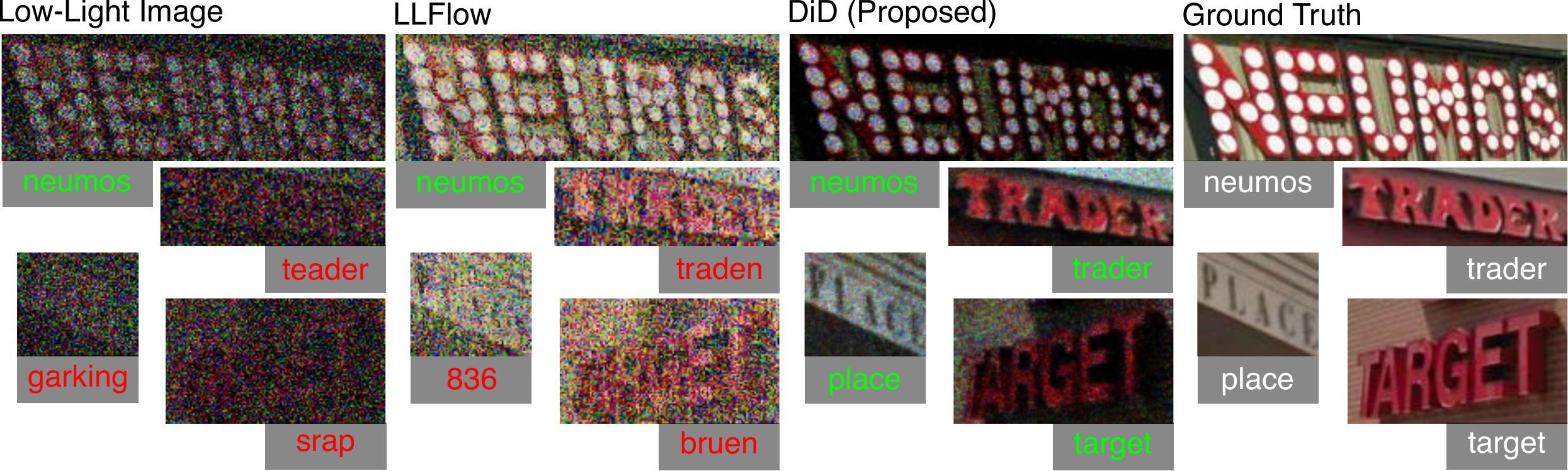}
    \captionof{figure}{Low-light image reconstruction methods often are not able to recover the fine detail necessary for high-level tasks. We propose a diffusion model to reconstruct not only suitable well-lit images, but also fine details required for text recognition. Here, we recover fine details around lettering, allowing downstream models to accurately identify text. Red and green signify incorrect and correct predictions, respectively. White text is ground truth.}
    \label{fig:teaser}
\end{center}
}]

%%%%%%%%% ABSTRACT
\begin{abstract}
% \vspace{-0.5cm}
Capturing images is a key part of automation for high-level tasks such as scene text recognition. 
Low-light conditions pose a challenge for high-level perception stacks, which are often optimized on well-lit, artifact-free images. 
Reconstruction methods for low-light images can produce well-lit counterparts, but typically at the cost of high-frequency details critical for downstream tasks. We propose Diffusion in the Dark (DiD), a diffusion model for low-light image reconstruction for text recognition. 
DiD provides qualitatively competitive reconstructions with that of state-of-the-art (SOTA), while preserving high-frequency details even in extremely noisy, dark conditions.
We demonstrate that DiD, without any task-specific optimization, can outperform SOTA low-light methods in low-light text recognition on real images, bolstering the potential of diffusion models to solve ill-posed inverse problems. 
Our code and pretrained models can be found on \url{https://ccnguyen.github.io/diffusion-in-the-dark/}.
\end{abstract}

%%%%%%%%% BODY TEXT

\section{Introduction}
Task automation has become ubiquitous. 
From the reading of license plates on highways to identifying groceries in a self-checkout line, automated tasks, powered by artificial intelligence, are everywhere and extremely reliant on visual cues, such as RGB images.
However, real-world imaging is subject to noisy conditions, optical blurs, and other aberrations that make downstream applications challenging. 
Notably, image post-processing pipelines used to improve the quality of these images are often designed to fulfill perceptual and aesthetic requirements, as decided by a human expert.
While these images may be useful for observation, said post-processing can fail to preserve high-frequency details, which may not be necessary for viewing pleasure but are critical for downstream applications, such as text recognition.

A particular challenge arises in low-light conditions.
Low-light images can have extremely low-photon counts, making it difficult to resolve the low signal-to-noise ratios~\cite{brooks2019unprocessing,wei2020physics}.
Convolutional neural networks (CNNs) have emerged as useful tools for low-light reconstruction~\cite{lore2017llnet,guo2020zero,wei2018deep,chen2018learning,maharjan2019improving}.
However, they are not very robust in low light as they fail to hallucinate details when there is very low signal to work from.
Generative models, on the other hand, have proven successful at recovering signals from low light, thanks to their ability to model a distribution of well-lit images.
These include generative adversarial networks (GANs)~\cite{jiang2021enlightengan} and normalizing flows~\cite{wang2022low}.
These methods are often designed to recover aesthetics, as seen in LLFlow~\cite{wang2022low} in Figure \ref{fig:teaser}.
We seek a method that not only recovers a well-lit image, but one that reconstructs high-frequency details useful for high-level tasks, such as text recognition. 
We focus on text recognition specifically because it requires fine details more so than other tasks, such as segmentation, as the goal is to predict entire words correctly.

Among the classes of generative models are diffusion models~\cite{song2019generative,sohl2015deep,ho2020denoising}, which iteratively denoise from random noise to reconstruct desired data samples. 
Compared to other generative models, diffusion models are stable in training and provide diversity in reconstructions, which allows a higher probability of reconstructing an optimal signal. 
Thus, we propose Diffusion in the Dark (DiD), a diffusion-based method for low-light image reconstruction. 
We train a diffusion model to reconstruct well-lit images that not only are aesthetically pleasing but also preserve fine-grain detail necessary for text recognition better than state-of-the-art (SOTA) low-light methods do. 
Specifically, we make the following contributions:
\begin{itemize}
\item We introduce a novel low-light reconstruction method for text recognition using a conditional diffusion model. DiD can reconstruct images at different resolutions, while training only on patches, reducing training time and computational cost.
\item We introduce key normalizations for training diffusion models on extremely dark or right-tailed data.
\item Through evaluations of baselines and ablation studies, we demonstrate that DiD provides the best reconstruction of low-light images for text recognition, without any task specific design, when compared to that of SOTA reconstruction methods, while not reducing the aesthetic quality. 
% \item \cindy{We show DiD performs well in reconstructing from unseen, real low-light scenes.}
% Through evaluations of baselines and ablation studies, we demonstrate that DiD provides better reconstruction of low-light images for text recognition, a high-level downstream task. Despite not outperforming the SOTA in image reconstruction in perceptual metrics, DiD provides competitive qualitative results and consistent preservation of high-frequency details in extremely dark, noisy conditions. \alex{I still think can make this more positive. For example "... we demonstrate that DiD provides better reconstruction of low-light images for text recognition, a high-level downstream task, while not qualitatively reducing the aesthetic quality perceived by humans. DiD provides competitive quantitative results with the existing SOTA in image reconstruction, and consistently preserves high-frequency details in extremely dark, noise conditions." Maybe we should do a user study lol.}
\end{itemize}
Without optimizing for a specific task, we demonstrate that DDPMs can reconstruct high-frequency detail better than exisiting generative models, and they show promise for other high-level tasks.
DiD provides competitive quantitative results with the SOTA and consistently preserves high-frequency details in extremely dark, noisy conditions.
We also show DiD performs well in reconstructing from unseen, real low-light scenes.

Diffusion models offer a new promising avenue for image reconstruction, one that is easy to train and can obtain better sample quality~\cite{dhariwal2021diffusion} over other generative models. 
It is vital to understand their potential and limitations in corner cases, such as low light.

% the potential and limitations of diffusion models for solving ill-posed inverse problems, while putting them to the test in corner cases, such as low light.
As more images are consumed by high-level perception stacks, we must examine how to better design reconstruction methods for complex tasks, and our work provides an encouraging step in that direction.

\section{Related Work}
\textbf{Low-light image enhancement.}
Classical low-light reconstruction methods include histogram equalization-based methods~\cite{abdullah2007dynamic,ibrahim2007brightness} and Retinex-based methods~\cite{fu2015probabilistic,li2018structure,park2017low,guo2016lime}.
The former performs a global transformation of an image using color histograms, while Retinex-based methods decompose light into reflectance and illuminance properties and use these as bases for reconstruction.
Burst averaging can also be used to mitigate noise in low-light scenarios~\cite{mildenhall2018burst,hasinoff2016burst,liu2014fast,liba2019handheld}, but these methods typically require extensive alignment procedures during post-processing to prevent ghosting artifacts and multiple photos. 
We opt to do single-image reconstruction.

Newer approaches use deep learning to not only advance aforementioned classical methods~\cite{liu2021retinex,wei2018deep,guo2020zero,wang2019rdgan}, but also bring new levels of robustness against extreme noise in low light.
% Guo et al.~\cite{guo2020zero} developed Zero-DCE, which estimates a set of image enhancing curves with each pixel learning its own curve to reconstruct well-lit images.
Zhang et al.~\cite{zhang2019kindling} use a network, KinD, to decouple illumination and reflectance. 
To fix non-uniform lighting artifacts in KinD, Zhang et al.~\cite{zhang2021beyond} developed KinD++, which uses multi-scale attention.
Wang et al.~\cite{wang2022low} developed LLFlow, using normalizing flows to capture the manifold of well-lit images by mapping them to a Gaussian distribution.
Given the difficulty of acquiring paired low-light/well-lit images, Jiang et al.~\cite{jiang2021enlightengan} propose an unsupervised GAN, using a global-local–focused discriminator and self-regularizing attention maps. 
% This method tends to leave some glare effects that can be suppressed using layer decomposition ~\cite{jin2022unsupervised}.
Zhou et al.~\cite{zhou2022lednet} perform low-light reconstruction in conjunction with deblurring to address both problems. 
Concurrent with our work, Yuan et al.~\cite{yuan2022learning} use conditional Denoising Diffusion Probabilistic Models (DDPMs) with stochastic corruptions during training to enhance night sky appearance on a small-scale dataset. 
Their method focuses on hallucinating plausible star appearances, while we focus on recovering exposures and white balancing for visually appealing, diverse scenes.

\textbf{Diffusion models.}
Diffusion models are a rising form of probabilistic generative models which can generate diverse, high-resolution images~\cite{dhariwal2021diffusion}.
Diffusion models take many different forms including DDPMs~\cite{ho2020denoising}, score-based generative modeling~\cite{song2019generative}, and stochastic differential equations~\cite{song2020score}.
They all follow similar processes: a forward process which gradually adds noise to clean samples drawn from a prior distribution and a reverse process which reverses the corruption process to recover plausible samples from noise.
Diffusion models have been successful at many challenging image-based tasks such as unconditional image generation~\cite{rombach2022high}, inpainting~\cite{saharia2022palette,lugmayr2022repaint,batzolis2021conditional,kawar2022denoising}, colorization~\cite{saharia2022palette,kawar2022denoising},  image segmentation~\cite{baranchuk2021label,amit2021segdiff}, and medical imaging~\cite{song2021solving,wolleb2022diffusion}. 
We refer the reader to a survey~\cite{croitoru2022diffusion} for more applications.
Diffusion models offer an attractive alternative to other generative models, such as GANs and variational autoencoders (VAEs), thanks to their stability in training and ability to learn strong priors~\cite{dhariwal2021diffusion,nichol2021glide}.
We focus on DDPM, which uses a U-Net~\cite{ronneberger2015u}, simplifying the need for task-specific, tedious architecture design~\cite{saharia2022palette}.

We highlight that generative models are known to generally perform \textit{worse} on traditional metrics such as PSNR/SSIM.
An L2 loss minimizes mean squared error (MSE), which conveniently maximizes PSNR.
However, probabilistic generative models optimize for learning a representative distribution rather than learning a deterministic solution~\cite{theis2015note,huszar2015not}, which would maximize MSE.
\textit{Thus, we are not optimizing for high PSNR/SSIM, nor do we expect that predictions from generative models provide the best PSNR/SSIM.}
It is well known that MSE, and therefore PSNR, cannot capture perceptual similarities~\cite{gupta2011modified,wang2003multiscale,wang2004image,wang2009mean,sharif2018suitability,ding2021comparison}.
Higher PSNR also does not necessarily correspond to greater photorealism~\cite{ledig2017photo,cheon2018generative}.
LPIPS~\cite{zhang2018unreasonable} and FID/KID~\cite{heusel2017gans,binkowski2018demystifying} are more representative metrics.
However, generative models are able to predict a wide range of exposure levels, and LPIPS is sensitive to different exposure levels, despite many exposure levels providing reasonable reconstructions.
See the supplement for more details.

\begin{figure*}
\begin{center}
\includegraphics[scale=0.58]{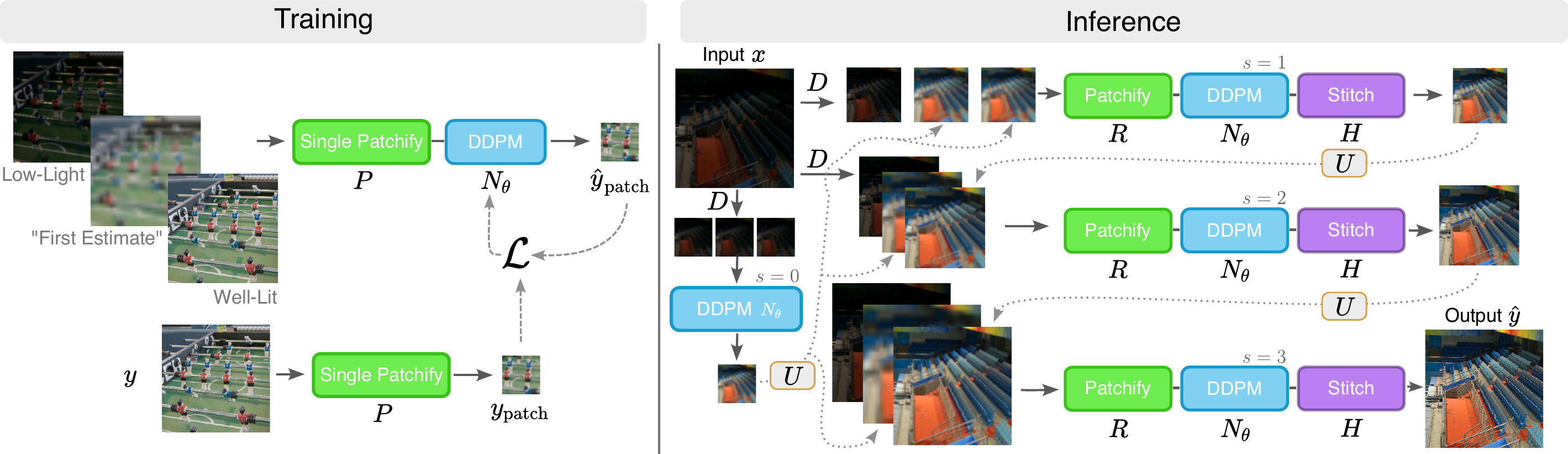}
\vspace{-0.5cm}
\end{center}
   \caption{\textbf{Overview of DiD pipeline.}  During training, we randomly crop $32 \times 32$ patches at multiple scales (noted with $s$)  using Single Patchify and concatenate the low-light patches, low-resolution well-lit patches, and high-resolution well-lit patches together as conditioning images to denoise and reconstruct well-lit patches. During inference, we use our trained DDPM network $N_\theta$ on 4 scales successively, each time to get well-lit patches at progressively better resolution. The prediction at each scale is upsampled using $U$ to use in the next scale.}
\label{fig:pipeline}
\vspace{-10pt}
\end{figure*}

\textbf{Domain transfer to text recognition.} 
Past works have demonstrated that perceptual metrics such as PSNR/SSIM are not indicative of success in downstream high-level tasks, such as image classification and segmentation~\cite{diamond2021dirty}. 
Task-specific imaging is an emerging paradigm, combating domain shift that high-level models experience on degraded images~\cite{wang2022robust, prakash2021gan,klinghoffer2022physics}.
Diamond et al.~\cite{diamond2021dirty} demonstrate that optimizing for perceptual metrics specifically can throw away details necessary for successful classification.

Modern scene text recognition (STR) methods~\cite{yao2014strokelets,liao2019scene,zhan2019esir,liao2019scene,qiao2020seed,yu2020towards,fang2022abinet++,bautista2022scene}, as noted in a recent survey~\cite{long2021scene}, only consider well-lit conditions.
Their performances tend to suffer in uneven or poor lighting.
Xue et al.~\cite{xue2020arbitrarily} combine spatial- and frequency-based features to enhance details for recognition of low-light text.
However, their method does not report very high precision or recall on standard well-lit text datasets.
Hsu et al.~\cite{hsu2022extremely} use a text-based loss, simulating low light from the ICDAR 2015 dataset~\cite{karatzas2015icdar}.
Liu et al.~\cite{liu2022list} perform text recognition using feature pyramids.
% Liu et al. also simulate low-light
% We demonstrate that with no additional training or optimization on text-based datasets, we perform better on low-light text recognition.
We demonstrate that, without any task-specific engineering, we reconstruct fine details to perform robustly in dark, noisy conditions using SOTA text recognition methods~\cite{fang2022abinet++,du2022svtr,he2022visual,baek2021if,nguyen2021dictionary}, such as PARSeq~\cite{bautista2022scene}.

% \textbf{Extreme Low-Light Imaging.}
% Most low-light enhancement works operate on ISP-processed images, which have nonlinear operations such as compression applied to the raw data.
% However, more extreme low-lit scenes may only produce some viable signal in 12-bit or 14-bit forms of raw data.
% Chen et al.~\cite{chen2018learning} operate on raw sensor data to replace the ISP and released the Seeing in the Dark (SID) dataset that contains scenes in nearly zero lux conditions (0.1 - 5 lux). 
% Maharjan et al.~\cite{maharjan2019improving} developed a residual network, as opposed to the original authors' U-Net, to reconstruct from SID. 
% Atoum et al. also developed a color-wise attention version of learning to enhance SID~\cite{atoum2020color}. 
% SID has also been significantly sped up to operate at 32 fps on a GPU, using optimized network structures~\cite{lamba2021restoring}.

% \textbf{Neural ISPs}
% Souza and Heidrich learn color rendition accuracy 

% CameraNet~\cite{liang2021cameranet}
% Eilertsen et al.~\cite{eilertsen2017hdr}

\section{Method}
We propose training a single DDPM to recover high-frequency details of full-resolution low-light image (Fig. \ref{fig:pipeline}).
Training a full-resolution diffusion model is extremely computationally demanding, requiring days of training on multiple GPUs.
Prior methods address this by using a cascading strategy, either training a single model in multiple phases~\cite{yuan2022learning} or training multiple models, each operating at a different resolution~\cite{ho2022cascaded,saharia2022image}.
We train a single model at multiple resolutions simultaneously, using a multi-scale patch-based approach.
We describe the key design choices to train a single model on multiple scales that allows us to train on a single GPU (Sec. \ref{sec:design}), the conditioning used in training (Sec. \ref{sec:train}), and inference process to successively predict larger resolutions to get our final reconstruction (Sec. \ref{sec:inference}).
We also describe our normalization scheme, which allows us to train on right-tailed data (Sec. \ref{sec:norm}).

\subsection{Background}
Diffusion models have different model families, including Variance Preserving (VP)~\cite{song2020score}, Variance Exploding (VE)~\cite{song2020score}, and Elucidating Diffusion Models (EDM)~\cite{karras2022elucidating}. 
We use the EDM formulation which includes applying a higher-order Runge-Kutta method for sampling, preconditioning, and improved loss function. 

Specifically, Karras et al. formulate their DDPM $\mathcal{D}(\bm{x}; \sigma)$, where $\bm{x}$ is the noisy image and $\sigma$ is the noise level, as a function that minimizes the expected MSE denoising error for samples drawn from the clean data distribution $\bm{p}_{\text{data}}$. 
The preconditioning, which uses a noise-level--independent skip-connection, allows $\mathcal{D}$ to estimate either the clean image $\bm{y}$ or noise $\bm{n}$. 
We choose to optimize a loss according to a prediction of $\bm{y}$, which can be expressed as

\begin{align}
\label{formula}
    \mathbb{E}_{\sigma, \bm{y}, \bm{n}}\left[\lambda(\sigma)||\mathcal{D}(\bm{y}+\bm{n}; \sigma) - \bm{y}||_2^2\right],
\end{align}

\noindent where $\bm{y} \sim p_{\text{data}}$ and $\bm{n} \sim \mathcal{N}(\bm{0}, \sigma^2 \bm{I})$.
This formulation allows us to use additional guiding losses on the predicted clean image $\bm{y}$.

\subsubsection{Design Choices and Architecture}
\label{sec:design}
We discuss two key design choices for our method.
The first is to operate on multiple scales. 
We found that training a $256 \times 256$ model to perform low-light reconstruction was too memory-intensive, so we opted to work on $32 \times 32$ patches.
However, decomposing a low-light image to $32 \times 32$ patches, running DDPM on each patch, and then stitching the patches together led to patch-to-patch inconsistencies in exposures and white balancing. 
See the supplement for examples.
Here, there is no constraint to enforce all patches to have the same appearance. 
Thus, we have multiple scales, each using the recovered exposure from the first inference step and, later in the inference process, other recovered exposures from previous steps as conditioning.

A second choice is to have a single network training on randomly selected scales rather than have a network for each scale. 
While this may have resolved some scale ambiguities in prediction, we found that training 4 models, one for each scale, would take 4 times as much memory or time to achieve the same number of training iterations and reconstruction quality as 1 model, as shown in our ablations (Sec. \ref{sec:ablations}). 

\subsubsection{Training Phase}
\label{sec:train}
Given low-lit/well-lit training pairs $\bm{x}, \bm{y} \in \mathcal{R}^{256 \times 256 \times 3}$, let $\mathcal{S}$ be a random variable following the discrete uniform distribution over the set $\{0, 1, 2, 3\}$.
In each training iteration, we sample a random scale $s \sim \mathcal{S}$ and use the function $\gamma(s) = 2^{s+5}, \gamma: s \rightarrow Y$, in which $Y$ is the set $\{32, 64, 128, 256\}$, to acquire a fixed resolution for scale $s$.
These values are set to be multiples of the smallest patch size to simplify the upsampling further on.
We use three conditioning inputs in each iteration:
\begin{itemize}
    \item Low-light condition $c_x$: The low-light measurement. This image provides the basis for reconstruction. For each scale, we will downsample the measurement to the corresponding operating resolution $\gamma(s)\times \gamma(s)$. 
    \item Well-lit condition $c_{y_1}$: The well-lit, but low-resolution prediction from the previous scale. This image provides an exposure level to condition from, which is closer to ground truth than $c_x$, but without well-lit high-frequency detail.
    \item First estimate condition $c_{y_2}$: The well-lit, but low-resolution prediction from $s=0$. This image provides a globally uniform exposure level on which to condition, further constraining the recovered exposure level. During training, if $s \neq 0$, $c_{y_2}$ is simulated by downsampling then upsampling the well-lit ground truth image.
\end{itemize}

Notably, if $s = 0$, we do not have a $c_{y_1}$ or $c_{y_2}$ from a previous scale to condition on, so we define our conditioning $x_{{\text{patch}}_i}$ using the low-light input $x_i$ and a bilinear downsampling operation $D(x, k)$ to reduce the low-light input to resolution $k \times k$. 
The conditioning input can be written as
\vspace{-12pt}
\begin{multline}
    x_{{\text{patch}}_i} = ([D(x_i, \gamma(0)), D(x_i, \gamma(0)), D(x_i, \gamma(0))].
\end{multline}
% \alex{I'm still a bit confused by this. At the lowest scale, you get xpatch and ypatch from xi, yi. So if I read eqn 8 right, you are downsampling the low light image, using that for all of the low-light condition cx AND the well-lit condition cy1 AND the first estiamte condition? Why would you use a low-lit image for the two well-lit conditionings? Do you mean you are using yi (coming from GT) here, like in eqn 9-12 for the higher scales?}

\noindent Note that $\gamma(0) \times \gamma(0)$ or $32 \times 32$ will be our fixed training resolution.
We define $P(x)$ as a random $32 \times 32$ cropping function (called ``Single Patchify''), $U(x,k)$ as an upsampling operation to bring $x$ to resolution $k\times k$, and our noise as $\eta \sim \mathcal{N}(0, \sigma^2)$. 
If $s > 0$, we define the training pairs as
\begin{align}
    c_x &= D(x_i, \gamma(s)),\\
    c_{y_1} &= U(D(y_i, \gamma(s-1)), \gamma(s)) + \eta,\\
    c_{y_2} &= U(D(y_0, \gamma(0)), \gamma(s)) + \eta, \\
    (x_{\text{patch}_i},y_{\text{patch}_i}) &= (P([c_x, c_{y_1}, c_{y_2}]_i), P(y_i)).
\end{align}
We add $\eta$ to better resemble the noisy predictions to later scales in the inference phase. 
We then apply the forward diffusion process of adding noise to $x_{{\text{patch}}_i}$ ~\cite{karras2022elucidating}. 
We pass the corrupted images and a noise channel $n \in \mathcal{R}^{32\times32\times3}$ to our denoising network $\Psi_\theta$ to produce a reconstruction:
\begin{align}
    \hat{y}_{\text{patch}_i} = \Psi_\theta ([x_{\text{patch}_i}, n]).
\end{align}
During training, we train on only patches of $32 \times 32$, but randomly select the scale of each image. 
This builds a robust model capable of reconstructing at multiple resolutions. 

We evaluate the loss $\mathcal{L}_{\text{DiD}}(\hat{y}_{\text{patch}_i}, y_{\text{patch}_i})$ to learn the weights $\theta$, using Eqn. (\ref{formula}).
Denoising alone was not sufficient for the challenging task of low-light reconstruction. 
We add additional losses to each denoising step on the predicted clean image, applying MSE and LPIPS~\cite{zhang2018unreasonable}. 
We chose MSE to help reconstruct better sharp details in the images and LPIPS~\cite{zhang2018unreasonable} to enhance appearance.
We find that these losses increase text recognition accuracy in addition to improving overall reconstruction.

\subsubsection{Inference Phase}
\label{sec:inference}
We follow a cascaded approach to regress our final image using a single model (Algorithm ~\ref{alg:inference}). 
We begin with the known low-light measurement $x_i$ and compose the conditioning inputs. 
We apply the reverse diffusion process, and use the diffusion prediction at the current scale as input to the next scale. 
We continue this until we compose our final $256 \times 256$ resolution well-lit image.
We observed that, even though the predictions are based on the same exposure instantiation under conditioning, exposure levels and white balancing still varied from patch-to-patch. 
To achieve full resolution consistency, we needed an additional step: Iterative Latent Variable Refinement (ILVR)~\cite{choi2021ilvr}.
ILVR guides the generative process by blending the high frequencies of the current denoised estimate with the noised, low-resolution version of the reference image. 
At each step of reverse denoising, we replaced the low-frequency details of the prediction at the current scale with low-frequency content extracted from our conditioning image: the low-resolution but well-lit reconstruction from the previous scale.
This conditioning does not require any additional training as it is only used during inference. 

\begin{algorithm}
\caption{Inference pipeline in DiD. $R(x)$ decomposes the images into $M$ $32\times 32$ patches (``Patchify'').
$H(x)$ stitches the $M$ reconstructed patches to a full resolution image. }
\label{alg:inference}
\begin{algorithmic}
\State $x^0_{{\text{patch}}} \gets [D(x_i, \gamma(0)), D(x_i, \gamma(0)),  D(x_i, \gamma(0))]$
\State $y^0_{{\text{patch}}} \gets \Psi_\theta ([x^0_{{\text{patch}}_i}, n])$\Comment{Enhance}
\For{$s = 0$, $k{+}{+}$, while $s < 4$}
    \State $c_x \gets D(x_i, \gamma(s))$\Comment{Low-light condition}
    \State $c_{y_1} \gets U(y^{k-1}_i, \gamma(s)))$\Comment{Well-lit condition}
    \State $c_{y_2} \gets U(y^{0}_i, \gamma(s)))$\Comment{First estimate condition}
    \State $x^s_{{\text{patches}}} \gets R([c_x, c_{y_1}, c_{y_2}])$\Comment{Patchify}
    \State $y^s_{\text{patches}} \gets \Psi_\theta ([x^s_{\text{patches}}, n])$ \Comment{Denoise with ILVR}
    \State $y^s_i \gets H(y^s_{\text{patches}})$\Comment{Stitch patches}
\EndFor
\end{algorithmic}
\end{algorithm}
% \vspace{-1cm}

\begin{figure}[t]
\begin{center}
   \includegraphics[scale=0.32]{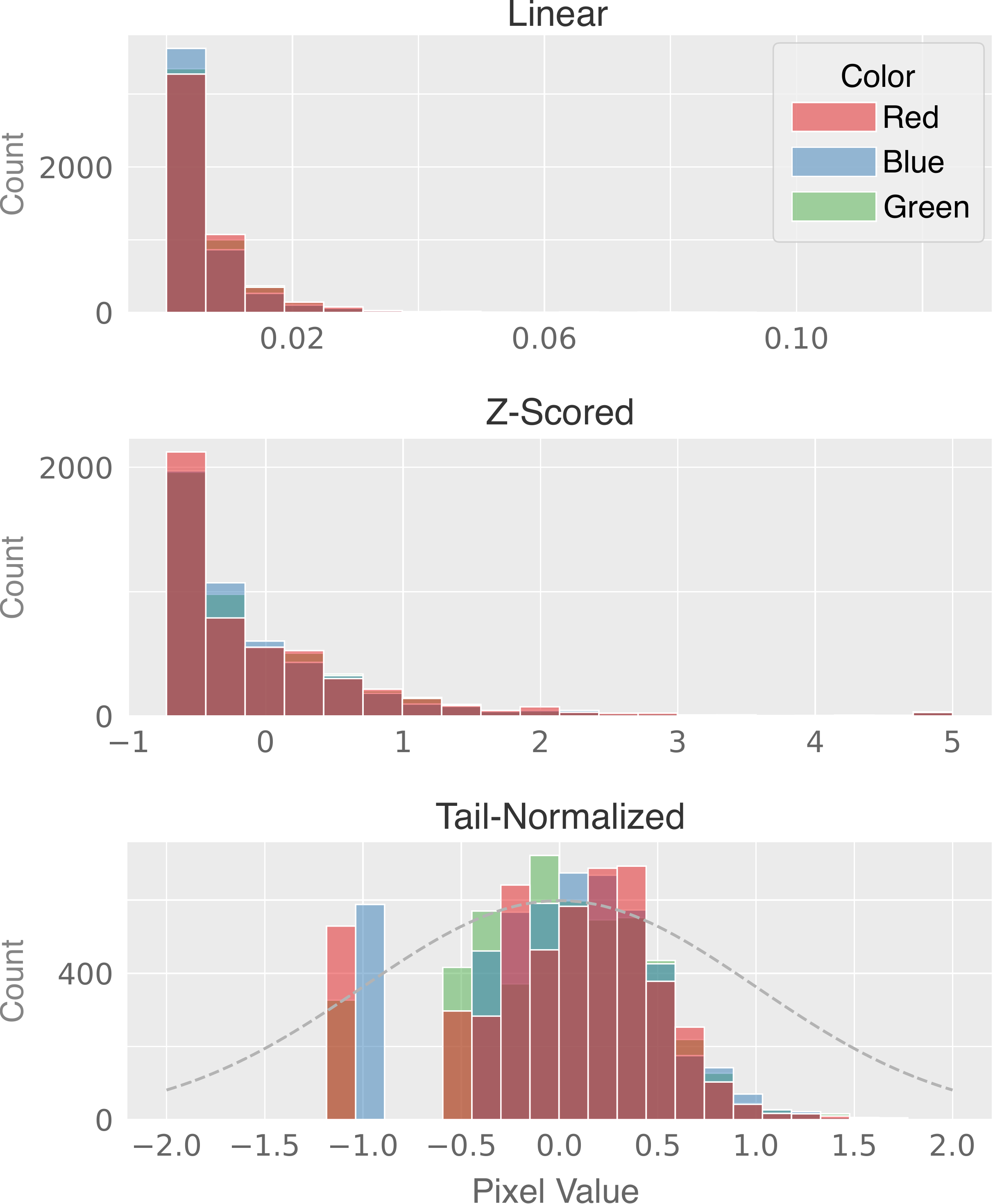}
\end{center}
\vspace{-0.6cm}
   \caption{\textbf{Color pixel distributions for low-light data.} Right-tailed data do not follow assumptions made in training a diffusion model. We normalize the data to the appropriate range for training. \textbf{Top:} Data distribution of a random selection of 30 images from the LOL training set~\cite{wei2018deep}. \textbf{Middle:} Data distribution of the same images after Z-scoring. The distribution is still right-tailed. \textbf{Bottom:} Data distribution of the same images from using tail-normalization (Sec. ~\ref{sec:norm}). The dotted line shows a true Gaussian distribution with $\mu = 0$ and $\sigma = 0.5$.}
\label{fig:data}
\vspace{-0.7cm}
\end{figure}

\subsection{Data Normalization}
\label{sec:norm}
It is useful to standardize data by centering and dividing with the sample standard deviation, so that each datum is represented in common units.
For example, DDPM~\cite{ho2020denoising} scales images to be within the range $[-1, 1]$. 
Given the right-tailed nature of low-light data (Fig. \ref{fig:data}), we cannot follow typical Z-scoring~\cite{zill2020advanced}. 
Diffusion models require picking a noise schedule at training, particularly a $\sigma_{min}$ and $\sigma_{max}$. 
The former is chosen such at that the lowest noise level is indistinguishable from images, and the latter is chosen such that the highest noise level is indistinguishable from white Gaussian noise. 
Since we are using $\sigma$ values designed for images, we need our images to be roughly in the same range to satisfy these conditions.
Thus, we normalize the data such that the distribution is between $[-1, 1]$ and follows a roughly Gaussian distribution with $\mu = 0$ and $\sigma = 0.5$. 
For right-tailed data, we found that taking the fourth root of the data, Z-scoring, and then dividing by two gave us a suitable distribution.
This normalization is critical as observed in our ablation studies (Sec. \ref{sec:ablations}).

\begin{figure*}
\begin{center}
\includegraphics[scale=0.42]{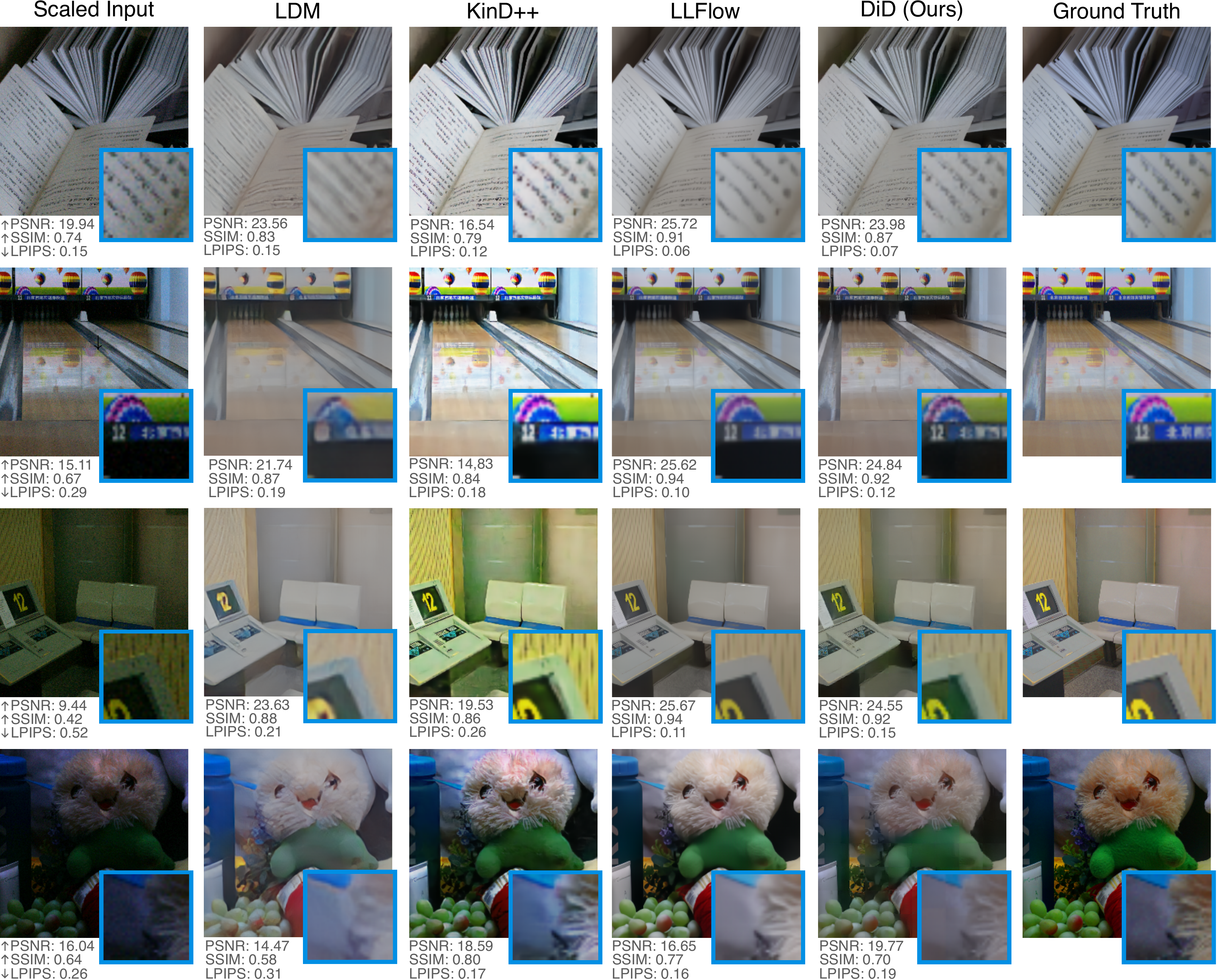}
\vspace{-0.5cm}
\end{center}
   \caption{\textbf{Qualitative results on the LOL test dataset.} We show reconstructions from LDM~\cite{rombach2022high} and the two best-performing low-light baselines, KinD++~\cite{zhang2021beyond} and LLFlow~\cite{wang2022low}. DiD reconstruction is on par with other low-light reconstruction methods while recovering more fine details such as handwriting and text on signs, such as the ``12'', with non-saturated appearances and reasonable exposure levels.}
\label{fig:qual1}
% \vspace{-0.5cm}
\end{figure*}

\section{Experiments}

\setlength{\tabcolsep}{2pt}

\begin{table}
\begin{center}
\caption{\textbf{Quantitative comparison of low-light enhancement methods on the LOL test dataset~\cite{wei2018deep}}. Diffusion models are separated from low-light--specific models.
We report the average inference time of a $256\times256$ image on an NVIDIA RTX 3090. $\lozenge$ indicates methods that did not closely match their reported performance.
We highlight the best and second best results using \textbf{bold} and \underline{underline}, respectively.}
\label{table:baselines}
\resizebox{\columnwidth}{!}{
% \begin{tabular}{l|cc}
% \hline\noalign{\smallskip}
% Method & Time (s) & \# Params \\
% \noalign{\smallskip}
% \hline
% \noalign{\smallskip}
% LDM~\cite{rombach2022high} & 9.809 & 404 M \\
% LLFlow~\cite{wang2022low} & 0.255 & 38.9 M\\
% DiD (Ours) & 6.635 & 55.7 M \\
% \hline
% \end{tabular}

\begin{tabular}{l|ccc|cc}
\hline\noalign{\smallskip}
Method & PSNR$\uparrow$ & SSIM$\uparrow$ & LPIPS$\downarrow$ & Time (s) & \# Params \\
\noalign{\smallskip}
\hline
\noalign{\smallskip}
Zero-DCE ~\cite{guo2020zero} & 14.67 & 0.68 & 0.35 & 0.06 & 79.4 K \\

LIME ~\cite{guo2016lime} $\lozenge$ & 14.22 & 0.65 & 0.37 & 1.10 & N/A\\

EnlightenGAN ~\cite{jiang2021enlightengan} & 16.84 & 0.76 & 0.33 & 0.08 & 8.6 M\\

RetinexNet ~\cite{wei2018deep} & 16.77 & 0.59 & 0.47 & 0.22 & 444.6 K\\

RUAS ~\cite{liu2021retinex} $\lozenge$ & 16.41 & 0.65 & 0.27 & 0.11 & 3.4 K \\

KinD ~\cite{zhang2019kindling} & 20.39 & 0.88 & \underline{0.16} & 0.33 & 8.0 M\\

KinD++ ~\cite{zhang2021beyond} & 21.73 & 0.89 & \underline{0.16} & 0.66 & 8.2 M\\

LLFlow ~\cite{wang2022low} & \textbf{24.94} & \textbf{0.91} & \textbf{0.12} & 0.26 & 38.9 M\\ 

\noalign{\smallskip}
\hline
\noalign{\smallskip}
DDRM ~\cite{kawar2022denoising} & 16.41 & 0.65 & 0.21 & 9.03 & 552.8 M\\ 
LDM ~\cite{rombach2022high} & 21.41	& 0.75 & 0.23 & 9.81 & 404.1 M\\
DiD (Ours) & \underline{23.97} & \underline{0.84} & \textbf{0.12} & 6.64 & 55.7 M \\
\hline
\end{tabular}

}
\end{center}
\vspace{-0.8cm}
\end{table}

\setlength{\tabcolsep}{1.4pt}

\begin{figure*}
\begin{center}
\includegraphics[scale=0.38]{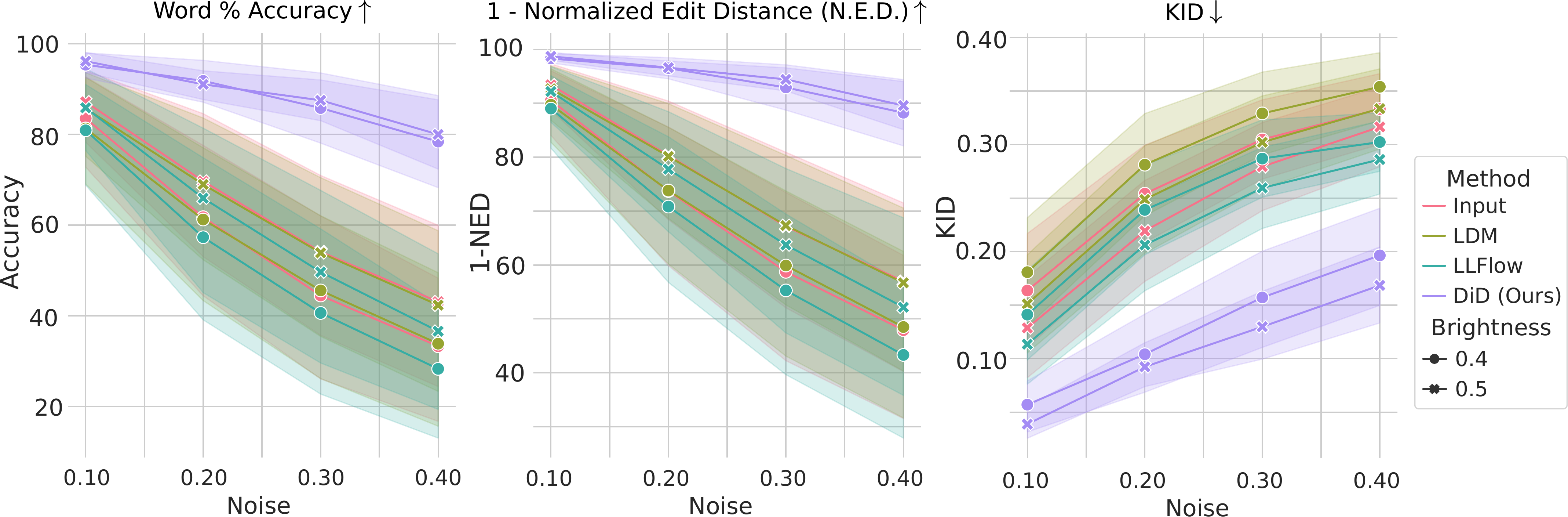}
\end{center}
   \caption{\textbf{Text recognition accuracy and reconstruction quality.} \textbf{Left and Center:} We plot word accuracy and 1-NED values at combinations of brightness and Poisson-Gaussian noise levels. Plotted points are the mean values of \textit{all} tested STR datasets. As brightness decreases and noise increases, other low-light reconstruction methods fail to recover enough detail to perform accurate text recognition, while DiD performs consistently well. \textbf{Right:} We display KID values for LLFlow and DiD, affirming that DiD provides reconstructions closer to the distribution of ground truth images.}
\label{fig:ocr}
\end{figure*}

\subsection{Implementation}
\vspace{-0.1cm}
We implement our framework with PyTorch on an NVIDIA Quadro RTX 8000. We use the ADAM optimizer~\cite{kingma2014adam} with a learning rate of $8 \times 10^{-4}$ and betas $[0.9, 0.999]$. 
Our patch-based scheme reduces training time and computational demand, using only 1 GPU to train within 3 days. 
We train batch sizes of 160 for 3000 iterations. 
Given that we are working with small datasets, we found this number of iterations to be suitable for convergence. 
We augment the data by adding either random Gaussian blur or sharpening, as well as scaling brightness and saturation.

\subsection{Dataset and Evaluation}
We train on the LOw-Light dataset (LOL)~\cite{wei2018deep}, which contains 485 training and 15 test low-light/well-lit pairs. 
There are limited low-light text datasets~\cite{xue2020arbitrarily,liu2022list}, but these unfortunately do not come with well-lit counterparts. 
We were interested in estimating not only accurate text, but also viable reconstructions from our method.
Thus, we opt to use a benchmark low-light dataset for our training and evaluation.

We report PSNR, SSIM, and LPIPS~\cite{zhang2018unreasonable}.
Due to the ill-posedness of low-light reconstruction, there are many optimal solutions that do not share the same white balance and exposure level as its ground truth, meaning we do not surpass SOTA on PSNR/SSIM, but do match LPIPS performance.
We are unable to provide KID~\cite{binkowski2018demystifying} or FID~\cite{heusel2017gans} scores on LOL given the limited test set.
However, for our text recognition task, we report KID for text reconstructions on much larger scene text datasets (Sec. \ref{sec:lowtext}).
We demonstrate that our method exceeds other low-light reconstruction methods in low-light text recognition in Word Accuracy and Normalized Edit Distance, the Levenshtein distance between words ~\cite{shi2017icdar2017,chng2019icdar2019,zhang2019icdar}.
We also demonstrate that our qualitative results are on par with SOTA performance for low-light reconstruction.

\subsection{Baselines}
To test our overall reconstruction quality, we compare against SOTA methods in low-light image enhancement, reporting results on the LOL test dataset.
We include results from two other diffusion-based models: Denoising Diffusion Restoration Models (DDRM)~\cite{kawar2022denoising} and LDMs~\cite{rombach2022high}. 
DDRM applies a pretrained denoising diffusion model to an inverse problem. 
We pretrain the DDRM to denoise noisy ImageNet~\cite{deng2009imagenet} images, and use the network to denoise a brightened version of the low-light images. 
LDMs use a pretrained VAE to encode images from pixel space to a latent space and trains a diffusion model in latent space.

For the non-deterministic models, we generate ten reconstructions for each image and pick the image that gives us highest PSNR. 
Although DiD does not have the best results numerically (Tab. ~\ref{table:baselines}), DiD performance is on par with that of SOTA, especially in LPIPS. 
DiD also performs the best of the diffusion-based models with significantly less trainable parameters and inference time.
% In low-light enhancement, PSNR/SSIM are not necessarily indicative of a useful reconstruction ~\cite{ledig2017photo}. 
% PSNR measures exact pixel correspondence with ground truth, while, realistically, there could be many ground truth images consistent with the observed low-light image. 
Despite not beating PSNR/SSIM, our generative model reconstructs high-frequency details better than SOTA low-light methods can (Fig. ~\ref{fig:qual1}).

\subsection{Ablations}
\label{sec:ablations}
To understand the contribution of each model component, we conduct several ablations (Tab. ~\ref{table:ablations}). 
% Here, we report the performance of a randomly selected prediction.
We refer the number of models trained and the number of scales used in each model as the \textbf{model-to-scale ratio}.
A 1:4 ratio means we train 1 model with 4 different resolutions, and a 2:1 ratio means we train 2 models, each corresponding to its own unique resolution. 
DiD, which uses a 1:4 ratio, outperforms all other ablations using a 1:4 models-to-scale ratios. 
We find that reducing the number of models or scales per model does not provide the level of conditioning information necessary for refined predictions. 
We also show that ILVR is critical for conditioning a prediction to be within a restricted exposure range.
See the supplement for details on model-to-scale ratios and more ablations.
\setlength{\tabcolsep}{1pt}
\begin{table}
\begin{center}
\caption{\textbf{Ablation studies.} \textbf{Models/scales} refers to the number of models and scales for each model. \textbf{Noise} refers to the addition of noise on the conditioning image. \textbf{LPIPS} refers to an additional LPIPS loss. \textbf{Data} refers to data normalization. \textbf{Cond.} refers to adding $c_{y_2}$ to the conditioning input. We highlight the best and second best results using \textbf{bold} and \underline{underline}, respectively.}
\resizebox{\columnwidth}{!}{
\begin{tabular}{lcccc|ccc}
\hline\noalign{\smallskip}
Models/scales & Noise & LPIPS & Data & Cond. & PSNR$\uparrow$ & SSIM$\uparrow$ & LPIPS$\downarrow$\\
\noalign{\smallskip}
\hline
\noalign{\smallskip}
1 : 4 & \redmark & \redmark & \redmark & \redmark & 16.26 & 0.57 & 0.48\\
1 : 4 & \redmark & \greenmark & \greenmark & \redmark & 19.56 & 0.74 & 0.35\\
1 : 4 & \greenmark & \greenmark & \redmark & \redmark & 16.94 & 0.63 & 0.46\\
1 : 4 & \greenmark & \greenmark & \greenmark & \redmark & 17.62 & 0.74 & \underline{0.31}\\
4 : 1 & \greenmark & \redmark & \greenmark & \redmark & \underline{19.63} & \underline{0.80} & \textbf{0.14}\\
1 : 2 & \greenmark & \greenmark & \greenmark & \redmark & 17.49 & 0.72 & 0.33\\
1 : 2 & \greenmark & \greenmark & \greenmark & \greenmark & 18.37 & 0.73 & 0.33\\
2 : 1 & \greenmark & \greenmark & \greenmark & \greenmark & 19.35 & 0.72 & \underline{0.31}\\
DiD (no ILVR) & \greenmark & \greenmark & \greenmark & \greenmark & 17.78 & 0.72 & 0.36\\
DiD & \greenmark & \greenmark & \greenmark & \greenmark & \textbf{21.00} & \textbf{0.82} & \textbf{0.14}\\
\hline
\end{tabular}

% \begin{tabular}{c|cccccc|ccc}
% \hline\noalign{\smallskip}
% Number & Models & Noise & LPIPS & Norm. & Cond. & ILVR & PSNR$\uparrow$ & SSIM$\uparrow$ & LPIPS$\downarrow$\\
% \noalign{\smallskip}
% \hline
% \noalign{\smallskip}
% (1) & 1 w/ 4 scales & \redmark & \redmark & \redmark & \redmark & \greenmark & 16.264 & 0.570 & 0.480\\
% (2) & 1 w/ 4 scales & \redmark & \greenmark & \greenmark & \redmark & \greenmark & 19.560 & 0.743 & 0.346\\
% (3) & 1 w/ 4 scales & \greenmark & \greenmark & \redmark & \redmark & \greenmark & 16.935 & 0.632 & 0.458\\
% (4) & 1 w/ 4 scales & \greenmark & \greenmark & \greenmark & \redmark & \greenmark & 17.619 & 0.741 & \underline{0.312}\\
% (5) & 4 w/ 1 scale each & \greenmark & \redmark & \greenmark & \redmark & \greenmark & \underline{19.631} & \underline{0.801} & \textbf{0.137}\\
% (6) & 1 w/ 2 scales & \greenmark & \greenmark & \greenmark & \redmark & \greenmark & 17.490 & 0.723 & 0.325\\
% (7) & 1 w/ 2 scales & \greenmark & \greenmark & \greenmark & \greenmark & \greenmark & 18.365 & 0.732 & 0.330\\
% (8) & 2 w/ 1 scale each & \greenmark & \greenmark & \greenmark & \greenmark & \greenmark & 19.348 & 0.722 & 0.313\\
% (9) & 4 w/ 1 scale each & \greenmark & \greenmark & \greenmark & \greenmark & \redmark & 17.778 & 0.721 & 0.364\\
% DiD & 1 w/ 4 scales & \greenmark & \greenmark & \greenmark & \greenmark & \greenmark & \textbf{20.998} & \textbf{0.815} & \textbf{0.137}\\
% \hline
% \end{tabular}

}
\label{table:ablations}
\end{center}
\vspace{-0.5cm}
\end{table}
\setlength{\tabcolsep}{1.4pt}

\subsection{Low-light Text Recognition}
\label{sec:lowtext}
We demonstrate our method's utility in low-light STR.
We evaluate on real scene text datasets: IIIT5k-Words (IIIT5k)~\cite{mishra2012scene}, ICDAR 2013 (IC13-1015)~\cite{karatzas2013icdar}, Street View Text (SVT)~\cite{wang2011end}, and SVT-Perspective (SVTP)~\cite{phan2013recognizing}.
We simulate capturing these images in low light by dimming its brightness and adding Poisson-Gaussian noise. More information on the noise model can be found in the supplement.
We found that more complex noise simulation is only relevant in the case of extremely low light~\cite{wei2020physics,wei2021physics,monakhova2022dancing}.
The noise level is more challenging than that of LOL (compare inputs from Figure \ref{fig:qual1} and Figure \ref{fig:teaser}).
We use a SOTA STR method, PARSeq~\cite{bautista2022scene}, to evaluate each low-light methods pretrained on LOL, on recovering text.

\begin{figure}
\begin{center}
\includegraphics[scale=1.]{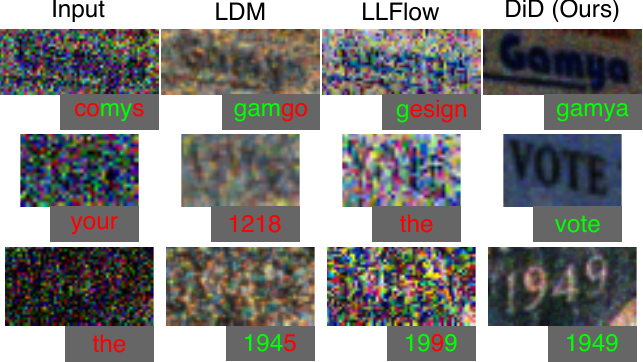}
\end{center}
\vspace{-0.4cm}
   \caption{\textbf{Comparing text recognition predictions.} We show samples from real scene text datasets and the reconstructions from LDM~\cite{rombach2022high}, LLFlow~\cite{wang2022low}, and DiD. 
   The input has been scaled down by a factor of 0.4 with Poisson-Gaussian noise.
   }
\vspace{-0.4cm}
\label{fig:wordqual}

\end{figure}

We report Word Accuracy (\% Acc) and Normalized Edit Distance (1 - NED) (Fig. \ref{fig:ocr}). 
DiD consistently performs well under extremely dark and noisy conditions, reporting $>75\%$ accuracy even in the most dark and noisy setting, while other methods begin to fail as conditions worsen.
We also present KID scores for LLFlow~\cite{wang2022low} and DiD in the supplement, which compares the distribution of the reconstructions to the distribution of the clean, well-lit ground truth STR images.
We reconstruct images that not only provide successful text recognition in dark conditions, but also better follow the distribution of ground truth images than other methods do (Fig. \ref{fig:wordqual}).

\subsection{Results on Other Datasets}
\begin{figure}
\begin{center}
\includegraphics[scale=0.3]{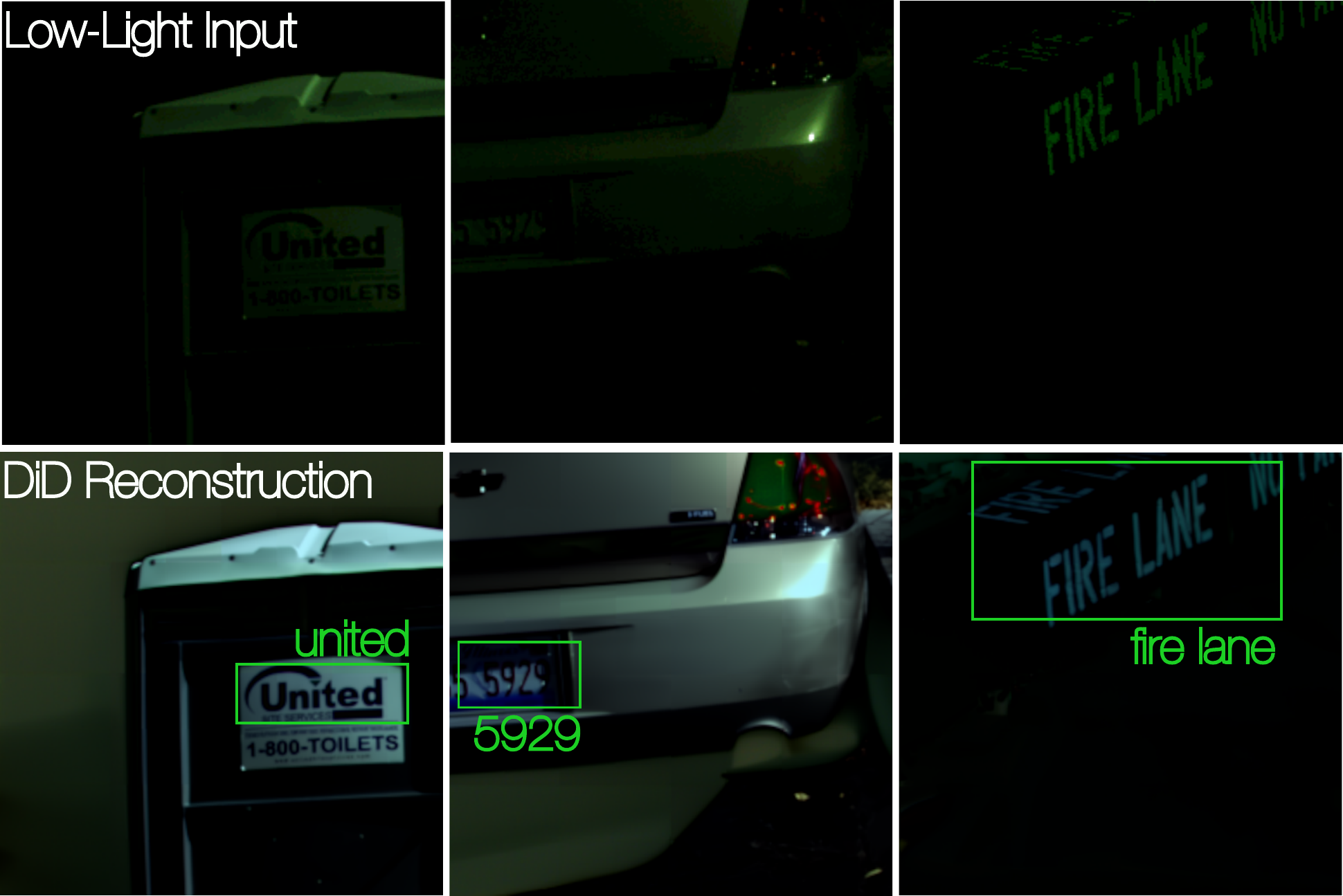}
\end{center}
\vspace{-0.4cm}
   \caption{
   \textbf{DiD reconstructs unseen, real low-light data.} Results from running PARSeq~\cite{bautista2022scene} on SID Sony low-light data~\cite{chen2018learning}. Green signifies correct text recognition. Low-light inputs are scaled for visualization.
   }
\vspace{-0.6cm}
\label{fig:sony}
\end{figure}

% We test our LOL-trained model on DARK FACE~\cite{wang2022unsupervised}, which consists of real-world low-light images.
% Our prediction [INSERT FIGURE] provides realistic reconstruction over other method trained on LOL, demonstrating that our method is robust against other real-world low-light datasets.
We tested our method on Seeing in the Dark (SID)~\cite{chen2018learning} Sony dataset, which consists of low-light RAW data with real noise. We trained our model on this RAW data, and found that its performance on the test set (PSNR: 23.98dB/SSIM: 0.78) was comparable to the original SID method (PSNR: 28.61dB/SSIM: 0.77). 
Our reconstruction allows for clear reconstruction of text in low light (Fig. ~\ref{fig:sony}). See our supplement for more dataset results.

\section{Discussion and Conclusions}
% summary
The future of automation depends on robust, high-level algorithms performing on images from a wide range of conditions. 
We propose a low-light reconstruction method using DDPMs which, without any specific task-level design, outperforms SOTA low-light reconstruction methods on low-light text recognition.

\textbf{Limitations and Future Work.}
Sampling from diffusion models requires multiple steps and may take prohibitively long in real-time scenarios. 
Fortunately, there is a fast growing family of methods to improve sampling time~\cite{lyu2022accelerating,kong2021fast,salimans2022progressive,watson2021learning} which can be applied.

 We use an autoregressive patch-based method which requires multiple passes through the network, and if image resolutions are not a multiple of the patch size, we have to interpolate the image to a larger resolution.
Future work would expand DiD for inference at all resolutions without interpolation, and LDMs~\cite{rombach2022high} and similar models which diffuse in latent space~\cite{avrahami2022blended,kwon2022diffusion} are a promising direction.

\textbf{Conclusion.}
As more tasks become automated, it is increasingly critical that reconstruction methods perform well in corner cases, such as dark and noisy conditions.
Our method provides sharp results, which align well with perceptual quality and especially downstream tasks operating on low-light images.

\small{
\noindent\textbf{Acknowledgements}
This project was in part supported by a PECASE from the ARO, by Samsung, and by the Stanford Institute for Human-Centered AI (HAI). We thank Yufei Wang for his help in running LLFlow. We thank Connor Lin and Nitish Padmanaban for their insights and comments on this work. 
}

%%%%%%%%% REFERENCES
{\small
\bibliographystyle{ieee_fullname}
\bibliography{egbib}

\begin{thebibliography}{100}\itemsep=-1pt

\bibitem{abdullah2007dynamic}
Mohammad Abdullah-Al-Wadud, Md~Hasanul Kabir, M~Ali~Akber Dewan, and Oksam Chae.
\newblock A dynamic histogram equalization for image contrast enhancement.
\newblock {\em IEEE Transactions on Consumer Electronics}, 53(2):593--600, 2007.

\bibitem{amit2021segdiff}
Tomer Amit, Eliya Nachmani, Tal Shaharbany, and Lior Wolf.
\newblock Segdiff: Image segmentation with diffusion probabilistic models.
\newblock {\em arXiv preprint arXiv:2112.00390}, 2021.

\bibitem{avrahami2022blended}
Omri Avrahami, Ohad Fried, and Dani Lischinski.
\newblock Blended latent diffusion.
\newblock {\em arXiv preprint arXiv:2206.02779}, 2022.

\bibitem{baek2021if}
Jeonghun Baek, Yusuke Matsui, and Kiyoharu Aizawa.
\newblock What if we only use real datasets for scene text recognition? toward scene text recognition with fewer labels.
\newblock In {\em Proceedings of the IEEE/CVF Conference on Computer Vision and Pattern Recognition}, pages 3113--3122, 2021.

\bibitem{baranchuk2021label}
Dmitry Baranchuk, Ivan Rubachev, Andrey Voynov, Valentin Khrulkov, and Artem Babenko.
\newblock Label-efficient semantic segmentation with diffusion models.
\newblock {\em arXiv preprint arXiv:2112.03126}, 2021.

\bibitem{batzolis2021conditional}
Georgios Batzolis, Jan Stanczuk, Carola-Bibiane Sch{\"o}nlieb, and Christian Etmann.
\newblock Conditional image generation with score-based diffusion models.
\newblock {\em arXiv preprint arXiv:2111.13606}, 2021.

\bibitem{bautista2022scene}
Darwin Bautista and Rowel Atienza.
\newblock Scene text recognition with permuted autoregressive sequence models.
\newblock In {\em Proceedings of the European Conference on Computer Vision (ECCV)}, pages 178--196. Springer, 2022.

\bibitem{binkowski2018demystifying}
Miko{\l}aj Bi{\'n}kowski, Danica~J Sutherland, Michael Arbel, and Arthur Gretton.
\newblock Demystifying mmd gans.
\newblock {\em arXiv preprint arXiv:1801.01401}, 2018.

\bibitem{brooks2019unprocessing}
Tim Brooks, Ben Mildenhall, Tianfan Xue, Jiawen Chen, Dillon Sharlet, and Jonathan~T Barron.
\newblock Unprocessing images for learned raw denoising.
\newblock In {\em Proceedings of the IEEE/CVF Conference on Computer Vision and Pattern Recognition}, pages 11036--11045, 2019.

\bibitem{chen2018learning}
Chen Chen, Qifeng Chen, Jia Xu, and Vladlen Koltun.
\newblock Learning to see in the dark.
\newblock In {\em Proceedings of the IEEE Conference on Computer Vision and Pattern Recognition}, pages 3291--3300, 2018.

\bibitem{cheon2018generative}
Manri Cheon, Jun-Hyuk Kim, Jun-Ho Choi, and Jong-Seok Lee.
\newblock Generative adversarial network-based image super-resolution using perceptual content losses.
\newblock In {\em Proceedings of the European Conference on Computer Vision (ECCV) Workshops}, pages 0--0, 2018.

\bibitem{chng2019icdar2019}
Chee~Kheng Chng, Yuliang Liu, Yipeng Sun, Chun~Chet Ng, Canjie Luo, Zihan Ni, ChuanMing Fang, Shuaitao Zhang, Junyu Han, Errui Ding, et~al.
\newblock Icdar2019 robust reading challenge on arbitrary-shaped text-rrc-art.
\newblock In {\em 2019 International Conference on Document Analysis and Recognition (ICDAR)}, pages 1571--1576. IEEE, 2019.

\bibitem{choi2021ilvr}
Jooyoung Choi, Sungwon Kim, Yonghyun Jeong, Youngjune Gwon, and Sungroh Yoon.
\newblock Ilvr: Conditioning method for denoising diffusion probabilistic models.
\newblock {\em arXiv preprint arXiv:2108.02938}, 2021.

\bibitem{croitoru2022diffusion}
Florinel-Alin Croitoru, Vlad Hondru, Radu~Tudor Ionescu, and Mubarak Shah.
\newblock Diffusion models in vision: A survey.
\newblock {\em arXiv preprint arXiv:2209.04747}, 2022.

\bibitem{dai2018dark}
Dengxin Dai and Luc Van~Gool.
\newblock Dark model adaptation: Semantic image segmentation from daytime to nighttime.
\newblock In {\em 2018 21st International Conference on Intelligent Transportation Systems (ITSC)}, pages 3819--3824. IEEE, 2018.

\bibitem{deng2009imagenet}
Jia Deng, Wei Dong, Richard Socher, Li-Jia Li, Kai Li, and Li Fei-Fei.
\newblock Imagenet: A large-scale hierarchical image database.
\newblock In {\em Proceedings of the IEEE Conference on Computer Vision and Pattern Recognition}, pages 248--255. Ieee, 2009.

\bibitem{dhariwal2021diffusion}
Prafulla Dhariwal and Alexander Nichol.
\newblock Diffusion models beat gans on image synthesis.
\newblock {\em Advances in Neural Information Processing Systems}, 34:8780--8794, 2021.

\bibitem{diamond2021dirty}
Steven Diamond, Vincent Sitzmann, Frank Julca-Aguilar, Stephen Boyd, Gordon Wetzstein, and Felix Heide.
\newblock Dirty pixels: Towards end-to-end image processing and perception.
\newblock {\em ACM Transactions on Graphics (TOG)}, 40(3):1--15, 2021.

\bibitem{ding2021comparison}
Keyan Ding, Kede Ma, Shiqi Wang, and Eero~P Simoncelli.
\newblock Comparison of full-reference image quality models for optimization of image processing systems.
\newblock {\em International Journal of Computer Vision}, 129:1258--1281, 2021.

\bibitem{du2022svtr}
Yongkun Du, Zhineng Chen, Caiyan Jia, Xiaoting Yin, Tianlun Zheng, Chenxia Li, Yuning Du, and Yu-Gang Jiang.
\newblock Svtr: Scene text recognition with a single visual model.
\newblock {\em arXiv preprint arXiv:2205.00159}, 2022.

\bibitem{fang2022abinet++}
Shancheng Fang, Zhendong Mao, Hongtao Xie, Yuxin Wang, Chenggang Yan, and Yongdong Zhang.
\newblock Abinet++: Autonomous, bidirectional and iterative language modeling for scene text spotting.
\newblock {\em IEEE Transactions on Pattern Analysis and Machine Intelligence}, 2022.

\bibitem{fu2015probabilistic}
Xueyang Fu, Yinghao Liao, Delu Zeng, Yue Huang, Xiao-Ping Zhang, and Xinghao Ding.
\newblock A probabilistic method for image enhancement with simultaneous illumination and reflectance estimation.
\newblock {\em IEEE Transactions on Image Processing}, 24(12):4965--4977, 2015.

\bibitem{guo2020zero}
Chunle Guo, Chongyi Li, Jichang Guo, Chen~Change Loy, Junhui Hou, Sam Kwong, and Runmin Cong.
\newblock Zero-reference deep curve estimation for low-light image enhancement.
\newblock In {\em Proceedings of the IEEE/CVF Conference on Computer Vision and Pattern Recognition}, pages 1780--1789, 2020.

\bibitem{guo2016lime}
Xiaojie Guo, Yu Li, and Haibin Ling.
\newblock Lime: Low-light image enhancement via illumination map estimation.
\newblock {\em IEEE Transactions on Image Processing}, 26(2):982--993, 2016.

\bibitem{gupta2011modified}
Prateek Gupta, Priyanka Srivastava, Satyam Bhardwaj, and Vikrant Bhateja.
\newblock A modified psnr metric based on hvs for quality assessment of color images.
\newblock In {\em 2011 International Conference on Communication and Industrial Application}, pages 1--4. IEEE, 2011.

\bibitem{hasinoff2016burst}
Samuel~W Hasinoff, Dillon Sharlet, Ryan Geiss, Andrew Adams, Jonathan~T Barron, Florian Kainz, Jiawen Chen, and Marc Levoy.
\newblock Burst photography for high dynamic range and low-light imaging on mobile cameras.
\newblock {\em ACM Transactions on Graphics (ToG)}, 35(6):1--12, 2016.

\bibitem{he2022visual}
Yue He, Chen Chen, Jing Zhang, Juhua Liu, Fengxiang He, Chaoyue Wang, and Bo Du.
\newblock Visual semantics allow for textual reasoning better in scene text recognition.
\newblock In {\em Proceedings of the AAAI Conference on Artificial Intelligence}, volume~36, pages 888--896, 2022.

\bibitem{heusel2017gans}
Martin Heusel, Hubert Ramsauer, Thomas Unterthiner, Bernhard Nessler, and Sepp Hochreiter.
\newblock Gans trained by a two time-scale update rule converge to a local nash equilibrium.
\newblock {\em Advances in Neural Information Processing Systems}, 30, 2017.

\bibitem{ho2020denoising}
Jonathan Ho, Ajay Jain, and Pieter Abbeel.
\newblock Denoising diffusion probabilistic models.
\newblock {\em Advances in Neural Information Processing Systems}, 33:6840--6851, 2020.

\bibitem{ho2022cascaded}
Jonathan Ho, Chitwan Saharia, William Chan, David~J Fleet, Mohammad Norouzi, and Tim Salimans.
\newblock Cascaded diffusion models for high fidelity image generation.
\newblock {\em J. Mach. Learn. Res.}, 23(47):1--33, 2022.

\bibitem{hsu2022extremely}
Po-Hao Hsu, Che-Tsung Lin, Chun~Chet Ng, Jie~Long Kew, Mei~Yih Tan, Shang-Hong Lai, Chee~Seng Chan, and Christopher Zach.
\newblock Extremely low-light image enhancement with scene text restoration.
\newblock In {\em 2022 26th International Conference on Pattern Recognition (ICPR)}, pages 317--323. IEEE, 2022.

\bibitem{huszar2015not}
Ferenc Husz{\'a}r.
\newblock How (not) to train your generative model: Scheduled sampling, likelihood, adversary?
\newblock {\em arXiv preprint arXiv:1511.05101}, 2015.

\bibitem{ibrahim2007brightness}
Haidi Ibrahim and Nicholas Sia~Pik Kong.
\newblock Brightness preserving dynamic histogram equalization for image contrast enhancement.
\newblock {\em IEEE Transactions on Consumer Electronics}, 53(4):1752--1758, 2007.

\bibitem{isola2017image}
Phillip Isola, Jun-Yan Zhu, Tinghui Zhou, and Alexei~A Efros.
\newblock Image-to-image translation with conditional adversarial networks.
\newblock In {\em Proceedings of the IEEE Conference on Computer Vision and Pattern Recognition}, pages 1125--1134, 2017.

\bibitem{jiang2021enlightengan}
Yifan Jiang, Xinyu Gong, Ding Liu, Yu Cheng, Chen Fang, Xiaohui Shen, Jianchao Yang, Pan Zhou, and Zhangyang Wang.
\newblock Enlightengan: Deep light enhancement without paired supervision.
\newblock {\em IEEE Transactions on Image Processing}, 30:2340--2349, 2021.

\bibitem{karatzas2015icdar}
Dimosthenis Karatzas, Lluis Gomez-Bigorda, Anguelos Nicolaou, Suman Ghosh, Andrew Bagdanov, Masakazu Iwamura, Jiri Matas, Lukas Neumann, Vijay~Ramaseshan Chandrasekhar, Shijian Lu, et~al.
\newblock Icdar 2015 competition on robust reading.
\newblock In {\em 2015 13th International Conference on Document Analysis and Recognition (ICDAR)}, pages 1156--1160. IEEE, 2015.

\bibitem{karatzas2013icdar}
Dimosthenis Karatzas, Faisal Shafait, Seiichi Uchida, Masakazu Iwamura, Lluis~Gomez i Bigorda, Sergi~Robles Mestre, Joan Mas, David~Fernandez Mota, Jon~Almazan Almazan, and Lluis~Pere De~Las~Heras.
\newblock Icdar 2013 robust reading competition.
\newblock In {\em 2013 12th International Conference on Document Analysis and Recognition (ICDAR)}, pages 1484--1493. IEEE, 2013.

\bibitem{karras2022elucidating}
Tero Karras, Miika Aittala, Timo Aila, and Samuli Laine.
\newblock Elucidating the design space of diffusion-based generative models.
\newblock {\em arXiv preprint arXiv:2206.00364}, 2022.

\bibitem{kawar2022denoising}
Bahjat Kawar, Michael Elad, Stefano Ermon, and Jiaming Song.
\newblock Denoising diffusion restoration models.
\newblock {\em arXiv preprint arXiv:2201.11793}, 2022.

\bibitem{kingma2014adam}
Diederik~P Kingma and Jimmy Ba.
\newblock Adam: A method for stochastic optimization.
\newblock {\em arXiv preprint arXiv:1412.6980}, 2014.

\bibitem{klinghoffer2022physics}
Tzofi Klinghoffer, Siddharth Somasundaram, Kushagra Tiwary, and Ramesh Raskar.
\newblock Physics vs. learned priors: Rethinking camera and algorithm design for task-specific imaging.
\newblock In {\em 2022 IEEE International Conference on Computational Photography (ICCP)}, pages 1--12. IEEE, 2022.

\bibitem{kong2021fast}
Zhifeng Kong and Wei Ping.
\newblock On fast sampling of diffusion probabilistic models.
\newblock {\em arXiv preprint arXiv:2106.00132}, 2021.

\bibitem{kwon2022diffusion}
Mingi Kwon, Jaeseok Jeong, and Youngjung Uh.
\newblock Diffusion models already have a semantic latent space.
\newblock {\em arXiv preprint arXiv:2210.10960}, 2022.

\bibitem{ledig2017photo}
Christian Ledig, Lucas Theis, Ferenc Husz{\'a}r, Jose Caballero, Andrew Cunningham, Alejandro Acosta, Andrew Aitken, Alykhan Tejani, Johannes Totz, Zehan Wang, et~al.
\newblock Photo-realistic single image super-resolution using a generative adversarial network.
\newblock In {\em Proceedings of the IEEE Conference on Computer Vision and Pattern Recognition}, pages 4681--4690, 2017.

\bibitem{lengyel2021zeroshot}
Attila Lengyel, Sourav Garg, Michael Milford, and Jan~C. van Gemert.
\newblock Zero-shot domain adaptation with a physics prior.
\newblock 2021.

\bibitem{li2018structure}
Mading Li, Jiaying Liu, Wenhan Yang, Xiaoyan Sun, and Zongming Guo.
\newblock Structure-revealing low-light image enhancement via robust retinex model.
\newblock {\em IEEE Transactions on Image Processing}, 27(6):2828--2841, 2018.

\bibitem{liao2019scene}
Minghui Liao, Jian Zhang, Zhaoyi Wan, Fengming Xie, Jiajun Liang, Pengyuan Lyu, Cong Yao, and Xiang Bai.
\newblock Scene text recognition from two-dimensional perspective.
\newblock In {\em Proceedings of the AAAI Conference on Artificial Intelligence}, volume~33, pages 8714--8721, 2019.

\bibitem{liba2019handheld}
Orly Liba, Kiran Murthy, Yun-Ta Tsai, Tim Brooks, Tianfan Xue, Nikhil Karnad, Qiurui He, Jonathan~T Barron, Dillon Sharlet, Ryan Geiss, et~al.
\newblock Handheld mobile photography in very low light.
\newblock {\em ACM Transactions on Graphics (TOG)}, 38(6):164--1, 2019.

\bibitem{liu2022list}
Hang Liu, Mengke Yuan, Tong Wang, Peiran Ren, and Dong-Ming Yan.
\newblock List: low illumination scene text detector with automatic feature enhancement.
\newblock {\em The Visual Computer}, 38(9-10):3231--3242, 2022.

\bibitem{liu2021retinex}
Risheng Liu, Long Ma, Jiaao Zhang, Xin Fan, and Zhongxuan Luo.
\newblock Retinex-inspired unrolling with cooperative prior architecture search for low-light image enhancement.
\newblock In {\em Proceedings of the IEEE/CVF Conference on Computer Vision and Pattern Recognition}, pages 10561--10570, 2021.

\bibitem{liu2014fast}
Ziwei Liu, Lu Yuan, Xiaoou Tang, Matt Uyttendaele, and Jian Sun.
\newblock Fast burst images denoising.
\newblock {\em ACM Transactions on Graphics (TOG)}, 33(6):1--9, 2014.

\bibitem{long2021scene}
Shangbang Long, Xin He, and Cong Yao.
\newblock Scene text detection and recognition: The deep learning era.
\newblock {\em International Journal of Computer Vision}, 129:161--184, 2021.

\bibitem{lore2017llnet}
Kin~Gwn Lore, Adedotun Akintayo, and Soumik Sarkar.
\newblock Llnet: A deep autoencoder approach to natural low-light image enhancement.
\newblock {\em Pattern Recognition}, 61:650--662, 2017.

\bibitem{lugmayr2022repaint}
Andreas Lugmayr, Martin Danelljan, Andres Romero, Fisher Yu, Radu Timofte, and Luc Van~Gool.
\newblock Repaint: Inpainting using denoising diffusion probabilistic models.
\newblock In {\em Proceedings of the IEEE/CVF Conference on Computer Vision and Pattern Recognition}, pages 11461--11471, 2022.

\bibitem{lv2021attention}
Feifan Lv, Yu Li, and Feng Lu.
\newblock Attention guided low-light image enhancement with a large scale low-light simulation dataset.
\newblock {\em International Journal of Computer Vision}, 129(7):2175--2193, 2021.

\bibitem{lyu2022accelerating}
Zhaoyang Lyu, Xudong Xu, Ceyuan Yang, Dahua Lin, and Bo Dai.
\newblock Accelerating diffusion models via early stop of the diffusion process.
\newblock {\em arXiv preprint arXiv:2205.12524}, 2022.

\bibitem{maharjan2019improving}
Paras Maharjan, Li Li, Zhu Li, Ning Xu, Chongyang Ma, and Yue Li.
\newblock Improving extreme low-light image denoising via residual learning.
\newblock In {\em 2019 IEEE International Conference on Multimedia and Expo (ICME)}, pages 916--921. IEEE, 2019.

\bibitem{mildenhall2018burst}
Ben Mildenhall, Jonathan~T Barron, Jiawen Chen, Dillon Sharlet, Ren Ng, and Robert Carroll.
\newblock Burst denoising with kernel prediction networks.
\newblock In {\em Proceedings of the IEEE Conference on Computer Vision and Pattern Recognition}, pages 2502--2510, 2018.

\bibitem{mishra2012scene}
Anand Mishra, Karteek Alahari, and CV Jawahar.
\newblock Scene text recognition using higher order language priors.
\newblock In {\em British Machine Vision Conference (BMVC)}. BMVA, 2012.

\bibitem{monakhova2022dancing}
Kristina Monakhova, Stephan~R Richter, Laura Waller, and Vladlen Koltun.
\newblock Dancing under the stars: video denoising in starlight.
\newblock In {\em Proc. CVPR IEEE}, pages 16241--16251, 2022.

\bibitem{nguyen2021dictionary}
Nguyen Nguyen, Thu Nguyen, Vinh Tran, Minh-Triet Tran, Thanh~Duc Ngo, Thien~Huu Nguyen, and Minh Hoai.
\newblock Dictionary-guided scene text recognition.
\newblock In {\em Proceedings of the IEEE/CVF Conference on Computer Vision and Pattern Recognition}, pages 7383--7392, 2021.

\bibitem{nichol2021glide}
Alex Nichol, Prafulla Dhariwal, Aditya Ramesh, Pranav Shyam, Pamela Mishkin, Bob McGrew, Ilya Sutskever, and Mark Chen.
\newblock Glide: Towards photorealistic image generation and editing with text-guided diffusion models.
\newblock {\em arXiv preprint arXiv:2112.10741}, 2021.

\bibitem{park2017low}
Seonhee Park, Soohwan Yu, Byeongho Moon, Seungyong Ko, and Joonki Paik.
\newblock Low-light image enhancement using variational optimization-based retinex model.
\newblock {\em IEEE Transactions on Consumer Electronics}, 63(2):178--184, 2017.

\bibitem{phan2013recognizing}
Trung~Quy Phan, Palaiahnakote Shivakumara, Shangxuan Tian, and Chew~Lim Tan.
\newblock Recognizing text with perspective distortion in natural scenes.
\newblock In {\em Proceedings of the IEEE International Conference on Computer Vision}, pages 569--576, 2013.

\bibitem{prakash2021gan}
Charan~D Prakash and Lina~J Karam.
\newblock It gan do better: Gan-based detection of objects on images with varying quality.
\newblock {\em IEEE Transactions on Image Processing}, 30:9220--9230, 2021.

\bibitem{qiao2020seed}
Zhi Qiao, Yu Zhou, Dongbao Yang, Yucan Zhou, and Weiping Wang.
\newblock Seed: Semantics enhanced encoder-decoder framework for scene text recognition.
\newblock In {\em Proceedings of the IEEE/CVF Conference on Computer Vision and Pattern Recognition}, pages 13528--13537, 2020.

\bibitem{rombach2022high}
Robin Rombach, Andreas Blattmann, Dominik Lorenz, Patrick Esser, and Bj{\"o}rn Ommer.
\newblock High-resolution image synthesis with latent diffusion models.
\newblock In {\em Proceedings of the IEEE/CVF Conference on Computer Vision and Pattern Recognition}, pages 10684--10695, 2022.

\bibitem{ronneberger2015u}
Olaf Ronneberger, Philipp Fischer, and Thomas Brox.
\newblock U-net: Convolutional networks for biomedical image segmentation.
\newblock In {\em International Conference on Medical Image Computing and Computer-Assisted Intervention}, pages 234--241. Springer, 2015.

\bibitem{saharia2022palette}
Chitwan Saharia, William Chan, Huiwen Chang, Chris Lee, Jonathan Ho, Tim Salimans, David Fleet, and Mohammad Norouzi.
\newblock Palette: Image-to-image diffusion models.
\newblock In {\em ACM SIGGRAPH 2022 Conference Proceedings}, pages 1--10, 2022.

\bibitem{saharia2022image}
Chitwan Saharia, Jonathan Ho, William Chan, Tim Salimans, David~J Fleet, and Mohammad Norouzi.
\newblock Image super-resolution via iterative refinement.
\newblock {\em IEEE Transactions on Pattern Analysis and Machine Intelligence}, 2022.

\bibitem{sakaridis2020map}
Christos Sakaridis, Dengxin Dai, and Luc Van~Gool.
\newblock Map-guided curriculum domain adaptation and uncertainty-aware evaluation for semantic nighttime image segmentation.
\newblock {\em IEEE Transactions on Pattern Analysis and Machine Intelligence}, 44(6):3139--3153, 2020.

\bibitem{sakaridis2021acdc}
Christos Sakaridis, Dengxin Dai, and Luc Van~Gool.
\newblock Acdc: The adverse conditions dataset with correspondences for semantic driving scene understanding.
\newblock In {\em Proceedings of the IEEE/CVF International Conference on Computer Vision}, pages 10765--10775, 2021.

\bibitem{salimans2022progressive}
Tim Salimans and Jonathan Ho.
\newblock Progressive distillation for fast sampling of diffusion models.
\newblock {\em arXiv preprint arXiv:2202.00512}, 2022.

\bibitem{sharif2018suitability}
Mahmood Sharif, Lujo Bauer, and Michael~K Reiter.
\newblock On the suitability of lp-norms for creating and preventing adversarial examples.
\newblock In {\em Proceedings of the IEEE Conference on Computer Vision and Pattern Recognition Workshops}, pages 1605--1613, 2018.

\bibitem{shi2017icdar2017}
Baoguang Shi, Cong Yao, Minghui Liao, Mingkun Yang, Pei Xu, Linyan Cui, Serge Belongie, Shijian Lu, and Xiang Bai.
\newblock Icdar2017 competition on reading chinese text in the wild (rctw-17).
\newblock In {\em 2017 14th IAPR International Conference on Document Analysis and Recognition (ICDAR)}, volume~1, pages 1429--1434. IEEE, 2017.

\bibitem{sohl2015deep}
Jascha Sohl-Dickstein, Eric Weiss, Niru Maheswaranathan, and Surya Ganguli.
\newblock Deep unsupervised learning using nonequilibrium thermodynamics.
\newblock In {\em International Conference on Machine Learning}, pages 2256--2265. PMLR, 2015.

\bibitem{song2019generative}
Yang Song and Stefano Ermon.
\newblock Generative modeling by estimating gradients of the data distribution.
\newblock {\em Advances in Neural Information Processing Systems}, 32, 2019.

\bibitem{song2021solving}
Yang Song, Liyue Shen, Lei Xing, and Stefano Ermon.
\newblock Solving inverse problems in medical imaging with score-based generative models.
\newblock {\em arXiv preprint arXiv:2111.08005}, 2021.

\bibitem{song2020score}
Yang Song, Jascha Sohl-Dickstein, Diederik~P Kingma, Abhishek Kumar, Stefano Ermon, and Ben Poole.
\newblock Score-based generative modeling through stochastic differential equations.
\newblock {\em arXiv preprint arXiv:2011.13456}, 2020.

\bibitem{theis2015note}
Lucas Theis, A{\"a}ron van~den Oord, and Matthias Bethge.
\newblock A note on the evaluation of generative models.
\newblock {\em arXiv preprint arXiv:1511.01844}, 2015.

\bibitem{wang2019rdgan}
Junyi Wang, Weimin Tan, Xuejing Niu, and Bo Yan.
\newblock Rdgan: Retinex decomposition based adversarial learning for low-light enhancement.
\newblock In {\em 2019 IEEE international conference on multimedia and expo (ICME)}, pages 1186--1191. IEEE, 2019.

\bibitem{wang2011end}
Kai Wang, Boris Babenko, and Serge Belongie.
\newblock End-to-end scene text recognition.
\newblock In {\em Proceedings of the IEEE International Conference on Computer Vision}, pages 1457--1464. IEEE, 2011.

\bibitem{wang2022robust}
Wen Wang, Jing Zhang, Wei Zhai, Yang Cao, and Dacheng Tao.
\newblock Robust object detection via adversarial novel style exploration.
\newblock {\em IEEE Transactions on Image Processing}, 31:1949--1962, 2022.

\bibitem{wang2022low}
Yufei Wang, Renjie Wan, Wenhan Yang, Haoliang Li, Lap-Pui Chau, and Alex Kot.
\newblock Low-light image enhancement with normalizing flow.
\newblock In {\em Proceedings of the AAAI Conference on Artificial Intelligence}, volume~36, pages 2604--2612, 2022.

\bibitem{wang2009mean}
Zhou Wang and Alan~C Bovik.
\newblock Mean squared error: Love it or leave it? a new look at signal fidelity measures.
\newblock {\em IEEE Signal Processing Magazine}, 26(1):98--117, 2009.

\bibitem{wang2004image}
Zhou Wang, Alan~C Bovik, Hamid~R Sheikh, and Eero~P Simoncelli.
\newblock Image quality assessment: from error visibility to structural similarity.
\newblock {\em IEEE Transactions on Image Processing}, 13(4):600--612, 2004.

\bibitem{wang2003multiscale}
Zhou Wang, Eero~P Simoncelli, and Alan~C Bovik.
\newblock Multiscale structural similarity for image quality assessment.
\newblock In {\em The Thrity-Seventh Asilomar Conference on Signals, Systems \& Computers, 2003}, volume~2, pages 1398--1402. Ieee, 2003.

\bibitem{watson2021learning}
Daniel Watson, Jonathan Ho, Mohammad Norouzi, and William Chan.
\newblock Learning to efficiently sample from diffusion probabilistic models.
\newblock {\em arXiv preprint arXiv:2106.03802}, 2021.

\bibitem{wei2018deep}
Chen Wei, Wenjing Wang, Wenhan Yang, and Jiaying Liu.
\newblock Deep retinex decomposition for low-light enhancement.
\newblock {\em arXiv preprint arXiv:1808.04560}, 2018.

\bibitem{Chen2018Retinex}
Chen Wei, Wenjing Wang, Wenhan Yang, and Jiaying Liu.
\newblock Deep retinex decomposition for low-light enhancement.
\newblock In {\em British Machine Vision Conference}, 2018.

\bibitem{wei2020physics}
Kaixuan Wei, Ying Fu, Jiaolong Yang, and Hua Huang.
\newblock A physics-based noise formation model for extreme low-light raw denoising.
\newblock In {\em Proceedings of the IEEE/CVF Conference on Computer Vision and Pattern Recognition}, pages 2758--2767, 2020.

\bibitem{wei2021physics}
Kaixuan Wei, Ying Fu, Yinqiang Zheng, and Jiaolong Yang.
\newblock Physics-based noise modeling for extreme low-light photography.
\newblock {\em IEEE TPAMI}, 44(11):8520--8537, 2021.

\bibitem{wolleb2022diffusion}
Julia Wolleb, Florentin Bieder, Robin Sandk{\"u}hler, and Philippe~C Cattin.
\newblock Diffusion models for medical anomaly detection.
\newblock {\em arXiv preprint arXiv:2203.04306}, 2022.

\bibitem{xu2020learning}
Ke Xu, Xin Yang, Baocai Yin, and Rynson~WH Lau.
\newblock Learning to restore low-light images via decomposition-and-enhancement.
\newblock In {\em Proceedings of the IEEE/CVF Conference on Computer Vision and Pattern Recognition}, pages 2281--2290, 2020.

\bibitem{xue2020arbitrarily}
Minglong Xue, Palaiahnakote Shivakumara, Chao Zhang, Yao Xiao, Tong Lu, Umapada Pal, Daniel Lopresti, and Zhibo Yang.
\newblock Arbitrarily-oriented text detection in low light natural scene images.
\newblock {\em IEEE Transactions on Multimedia}, 23:2706--2720, 2020.

\bibitem{yao2014strokelets}
Cong Yao, Xiang Bai, Baoguang Shi, and Wenyu Liu.
\newblock Strokelets: A learned multi-scale representation for scene text recognition.
\newblock In {\em Proceedings of the IEEE Conference on Computer Vision and Pattern Recognition}, pages 4042--4049, 2014.

\bibitem{yu2020towards}
Deli Yu, Xuan Li, Chengquan Zhang, Tao Liu, Junyu Han, Jingtuo Liu, and Errui Ding.
\newblock Towards accurate scene text recognition with semantic reasoning networks.
\newblock In {\em Proceedings of the IEEE/CVF Conference on Computer Vision and Pattern Recognition}, pages 12113--12122, 2020.

\bibitem{yuan2022learning}
Yu Yuan, Jiaqi Wu, Lindong Wang, Zhongliang Jing, Henry Leung, Shuyuan Zhu, and Han Pan.
\newblock Learning to kindle the starlight.
\newblock {\em arXiv preprint arXiv:2211.09206}, 2022.

\bibitem{zhan2019esir}
Fangneng Zhan and Shijian Lu.
\newblock Esir: End-to-end scene text recognition via iterative image rectification.
\newblock In {\em Proceedings of the IEEE/CVF Conference on Computer Vision and Pattern Recognition}, pages 2059--2068, 2019.

\bibitem{zhang2018unreasonable}
Richard Zhang, Phillip Isola, Alexei~A Efros, Eli Shechtman, and Oliver Wang.
\newblock The unreasonable effectiveness of deep features as a perceptual metric.
\newblock In {\em Proceedings of the IEEE Conference on Computer Vision and Pattern Recognition}, pages 586--595, 2018.

\bibitem{zhang2019icdar}
Rui Zhang, Yongsheng Zhou, Qianyi Jiang, Qi Song, Nan Li, Kai Zhou, Lei Wang, Dong Wang, Minghui Liao, Mingkun Yang, et~al.
\newblock Icdar 2019 robust reading challenge on reading chinese text on signboard.
\newblock In {\em 2019 International Conference on Document Analysis and Recognition (ICDAR)}, pages 1577--1581. IEEE, 2019.

\bibitem{zhang2021better}
Yu Zhang, Xiaoguang Di, Bin Zhang, Ruihang Ji, and Chunhui Wang.
\newblock Better than reference in low-light image enhancement: conditional re-enhancement network.
\newblock {\em IEEE Transactions on Image Processing}, 31:759--772, 2021.

\bibitem{zhang2021beyond}
Yonghua Zhang, Xiaojie Guo, Jiayi Ma, Wei Liu, and Jiawan Zhang.
\newblock Beyond brightening low-light images.
\newblock {\em International Journal of Computer Vision}, 129(4):1013--1037, 2021.

\bibitem{zhang2019kindling}
Yonghua Zhang, Jiawan Zhang, and Xiaojie Guo.
\newblock Kindling the darkness: A practical low-light image enhancer.
\newblock In {\em Proceedings of the 27th ACM International Conference on Multimedia}, pages 1632--1640, 2019.

\bibitem{zheng2022low}
Shen Zheng, Yiling Ma, Jinqian Pan, Changjie Lu, and Gaurav Gupta.
\newblock Low-light image and video enhancement: A comprehensive survey and beyond.
\newblock {\em arXiv preprint arXiv:2212.10772}, 2022.

\bibitem{zhou2022lednet}
Shangchen Zhou, Chongyi Li, and Chen Change~Loy.
\newblock Lednet: Joint low-light enhancement and deblurring in the dark.
\newblock In {\em European Conference on Computer Vision}, pages 573--589. Springer, 2022.

\bibitem{zhu2017unpaired}
Jun-Yan Zhu, Taesung Park, Phillip Isola, and Alexei~A Efros.
\newblock Unpaired image-to-image translation using cycle-consistent adversarial networks.
\newblock In {\em Proceedings of the IEEE International Conference on Computer Vision}, pages 2223--2232, 2017.

\bibitem{zill2020advanced}
Dennis~G Zill.
\newblock {\em Advanced engineering mathematics}.
\newblock Jones \& Bartlett Learning, 2020.

\end{thebibliography}
}

\clearpage
\newpage

\section*{Supplementary Material}
% \appendix
% \renewcommand{\thesection}{\Alph{section}}% Adjust section printing (from here onward)
\makeatletter
\renewcommand \thesection{S\@arabic\c@section}
\renewcommand\thetable{S\@arabic\c@table}
\renewcommand \thefigure{S\@arabic\c@figure}
\makeatother

\setcounter{section}{0}

\counterwithin{figure}{section}
\counterwithin{table}{section}
% \renewcommand\thefigure{\thesection\arabic{figure}}
% \renewcommand\thetable{\thesection\arabic{table}}
% \maketitle
\section{Teaser Figure}
Images used in the teaser figure are from the SVTP dataset at brightness level 0.4 and random Poisson-Gaussian noise level 0.25.

\section{Architecture}
\subsection{Network design}
One can get reasonable recovery of low-light signal by scaling up low-light images.
However, in scaling up low-light images, the noise is also amplified, and this is apparent in the scaled inputs visualized in Figure 4 of the main paper.
We wanted to reconstruct fine details without noise amplification from a single image, so we opt to use a generative model.

We tested ADM~\cite{dhariwal2021diffusion} as an alternative to DDPM and found instabilities in training.
We also tested a training patch resolution of $64 \times 64$ and found that it worked comparably, with slightly longer training times. 
We choose not to train a GAN such as CycleGAN~\cite{zhu2017unpaired} or Pix2Pix~\cite{isola2017image} because there is no large paired dataset of low-light/well-lit pairs, which would make training a GAN especially unstable.

\subsection{Network conditioning}
% Diffusion models rely on different methods for high resolution inference, including cascading models~\cite{ho2022cascaded,nichol2021glide,ramesh2022hierarchical}, large networks~\cite{dhariwal2021diffusion,song2020score}, or hybrid network approaches~\cite{preechakul2022diffusion,rombach2022high,vahdat2021score}.
% We follow the cascading approach using one model

We use a U-Net~\cite{ronneberger2015u} as the base of our DDPM. We use the U-Net as implemented in ~\cite{song2020score}, which consists of 3 ``down-blocks'' and 3 ``up-blocks'' with skip connections between them.
The network uses positional embeddings, 4 residual blocks per resolution, and per-resolution multipliers of [2,2,2]. 
The network has a base 128 number of channels, and a dropout factor of $0.10$.
We use the weighting of the L2 loss as prescribed by Karras et al.~\cite{karras2022elucidating} and apply a fixed scalar weight to the perceptual component (LPIPS ~\cite{zhang2018unreasonable}) of our custom loss function.

\begin{figure}[t]
\begin{center}
   \includegraphics[width=\linewidth]{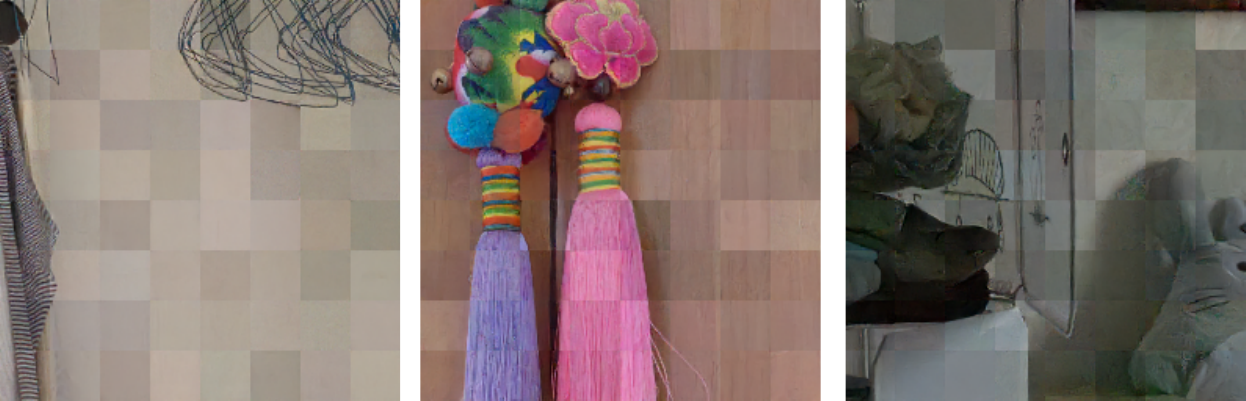}
\end{center}
% \vspace{-0.6cm}
   \caption{\textbf{Inference without exposure and white balancing constraints.} Reconstructions show patch-to-patch inconsistencies in exposure levels and white balancing if we perform inference on individual patches and stitch them together or we do not perform additional ILVR conditioning in DiD. }
\label{fig:patchy}
% \vspace{-0.6cm}
\end{figure}

We show examples of patch-to-patch inconsistencies observed without proper conditioning in Figure \ref{fig:patchy}. 
Using a multi-scale approach with ILVR~\cite{choi2021ilvr}, we can mitigate these issues to reconstruct a coherent image.
Using ILVR~\cite{choi2021ilvr} at every denoising step led to blurring, but applying ILVR to $6$ of the 18 steps was sufficient. 
For our loss, we empirically found that $\lambda = 5$ worked well. 
Before training, we apply EDM~\cite{karras2022elucidating} preconditioning.

\section{Training and Inference}
\subsection{Low-light datasets}
It is challenging to find large real low-light training datasets.
Multiple works have demonstrated accurate noise modeling for low-light~\cite{wei2020physics,brooks2019unprocessing}, but it remains difficult to model the loss of scene content and color in dim lighting. 
We opt to use the LOL dataset because it remains one of the most popular choices for low-light training~\cite{zheng2022low}, allowing for easier comparison against SOTA.

\subsection{Data preprocessing}
% An alternative method to performing high quality low-light text enhancement would be to use a generic low-light image enhancement method and then applying a denoising method. However, 
For tail-normalization, the exact root number and division number may vary from dataset to dataset. We find that our choice of fourth root and dividing by two after z-scoring was suitable for the LOL~\cite{wei2018deep}, Seeing in the Dark~\cite{chen2018learning}, and a modified Seeing in the Dark~\cite{xu2020learning} dataset.

\begin{figure}[t]
\begin{center}
   \includegraphics[width=\linewidth]{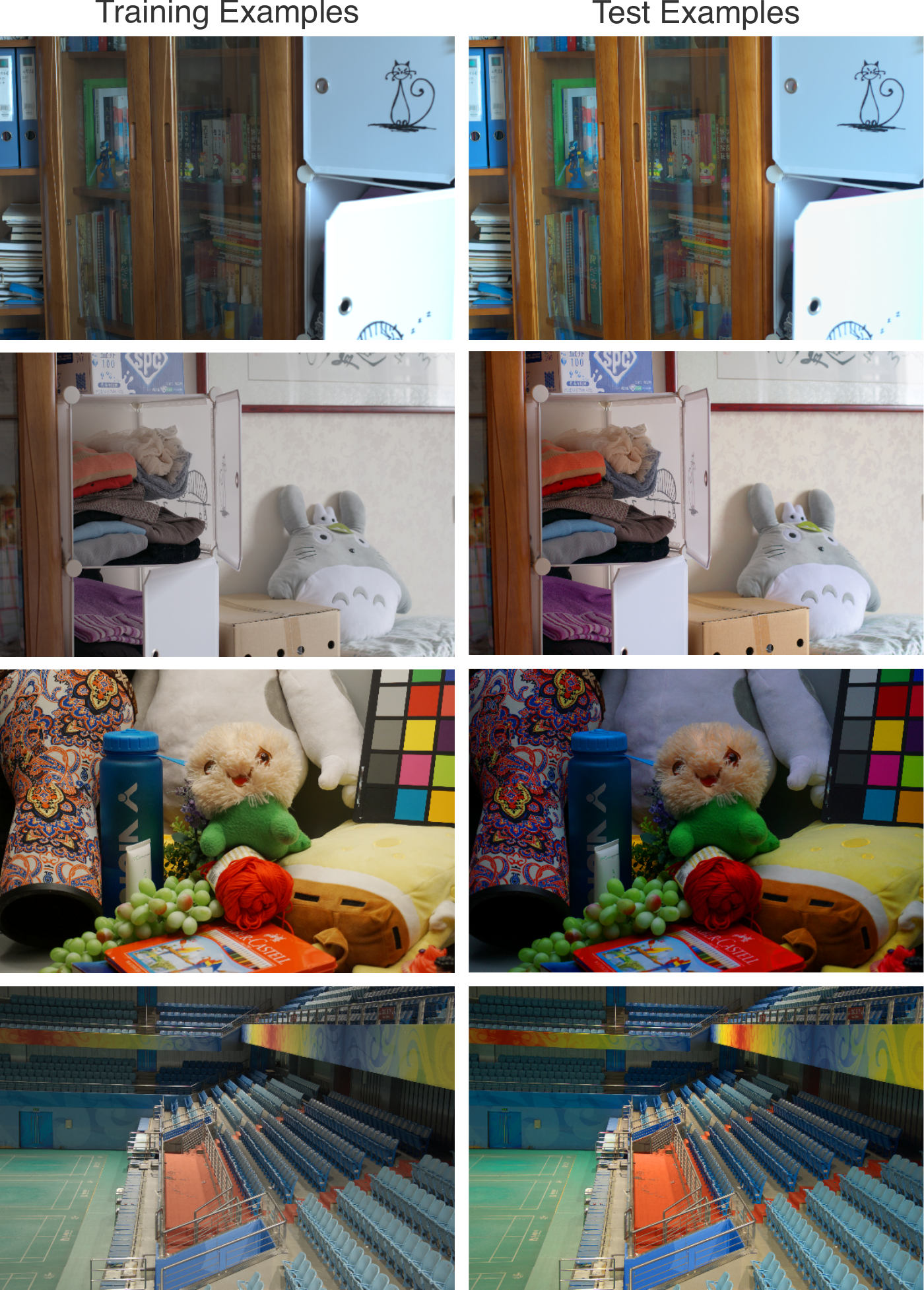}
\end{center}
% \vspace{-0.6cm}
   \caption{\textbf{LOL dataset inconsistencies.} There is overlap in scenes between the LOL training and test datasets, and the well-lit test images are significantly more contrasted than the well-lit training set images. }
\label{fig:lol_data}
% \vspace{-0.6cm}
\end{figure}

Among low-light datasets, LOL is the most popular to train and test on after custom datasets as found in a survey of low-light reconstruction methods~\cite{zheng2022low}.
However, there is significant overlap in scenes from the train and test set, and for unknown reasons, the test set ground truths have their color contrast raised, as seen in Figure \ref{fig:lol_data}.
This contrast raise makes it challenging to get an accurate sense of performance. 
We report quantitative results on the LOL test set for comparison sake, but believe that our image quality is reflective of realistic coloring as shown in the training set.

\begin{figure}[t]
\begin{center}
   \includegraphics[width=\linewidth]{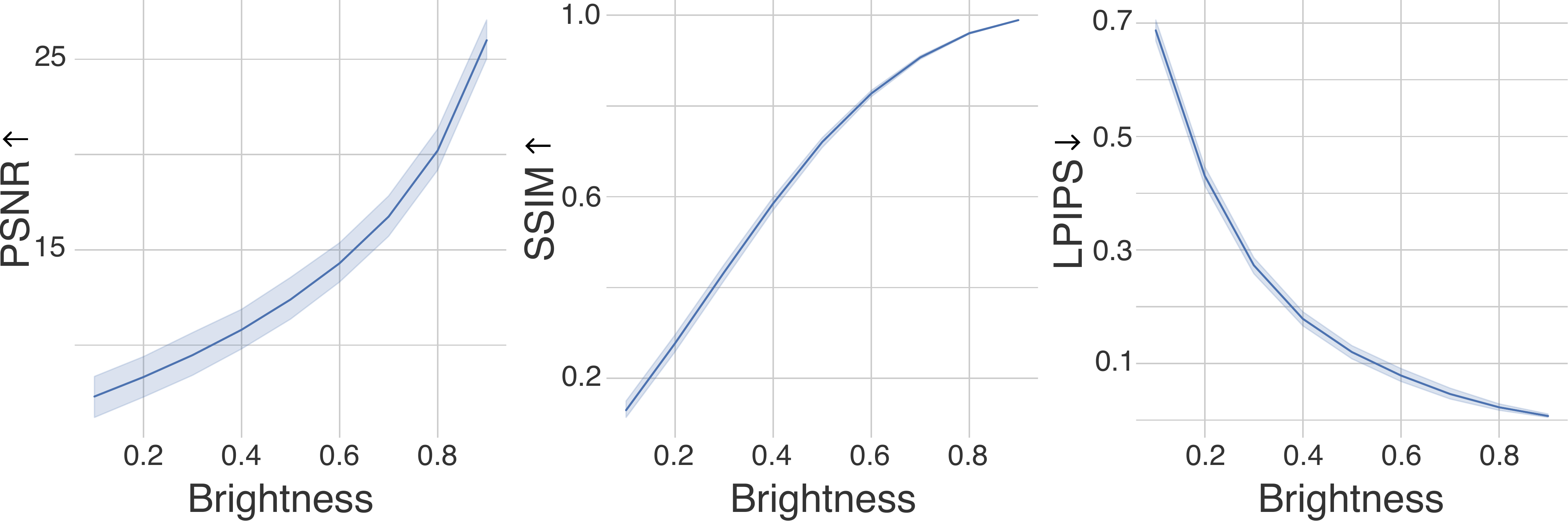}
\end{center}
% \vspace{-0.35cm}
   \caption{\textbf{Metrics are sensitive to exposure levels.} There is a significant drop in performance once exposure levels begin to change, even though content is the same. The brightness represented here is the scaled V value in the HSV-converted image.}
\label{fig:metrics}
% \vspace{-0.6cm}
\end{figure}

For LOL, we perform preprocessing before training. 
First, we center crop the image to be $256 \times 256$.
We then convert the image from sRGB space to linear space. 
We perform data normalization using a mean and standard deviation found in linear space on a random sample of 30 images. 
After tail-normalizing our data, we then train with images in the range $[-1, 1]$. 
Upon inference, we unnormalize the data and convert the image from linear space back to sRGB space for visualization. 
We compute all metrics in sRGB space.
We note that metrics are sensitive to different exposure levels despite having the same content. 
We show this sensitivity by scaling down the brightness of the LOL test dataset and computing PSNR, SSIM, and LPIPS across different brightness levels (Fig. \ref{fig:metrics}).

\subsection{Inference details}
For inference, we apply 18 sampling steps. 
For $s=0$, we do not apply ILVR. 
For $s=1,2,3$, we apply ILVR to the first 6 steps, using the low-frequency content from the previous scale's prediction. 
To filter low-frequency content, we downsample and upsample respectively, using bilinear interpolation with anti-aliasing. 
We also tested numerous different filters (Lanczos, cubic, and nearest) and found no significant difference in performance.
We tested using more inference steps ($N=100$) and found only minor changes in performance (PSNR: \textcolor{green}{$+0.120$}, SSIM: \textcolor{red}{$-0.002$}, LPIPS: \textcolor{green}{$-0.011$}). 

% \subsection{Computational and memory demand}
% We sought for a method that could be easily trained on a single GPU. 
% As seen in Table \ref{table:compute}, our method requires significantly less trainable parameters than LDM.
% Our time for inference is longer than LLFlow, but faster sampling methods for DDPMs is an area of active research out of the scope of this paper.

% \subsection{Training on RAW}
% We also conducted experiments on the well-known Seeing in the Dark (SID) dataset~\cite{chen2018learning}.
% The dataset consists of RAW data from real-life captures of indoor and outdoor scenes, primarily at nighttime, from a Sony $\alpha$7S II and Fujifilm X-T2 camera.
% We utilize the pre-processed version of the Sony subset of the dataset from Xu et al.~\cite{xu2020learning}.
% The additional pre-processing includes demosaicing, exposure compensation, white balancing, and delinearization of RAW data. 
% This version of the dataset includes 4,198 training pairs and 1,196 pairs for testing. 
% We find that our method is able to perform suitably when trained on this more challenging dataset as well. 

\section{Experiments}
\subsection{Baseline methods}
LLFlow~\cite{wang2022low} is non-deterministic in theory and deterministic in practice. 
The method uses a fixed latent feature, which leads to a deterministic result. 
By changing the latent feature, one can get different results due to the one-to-one mapping of normalizing flows. 
However, because the results from different latents are not perceptually obvious, we maintain a fixed latent feature. 
This choice may differ in cases where there is more training data.
For DDRM~\cite{kawar2022denoising}, we scale up the brightness of LOL images by a factor of 6 to produce a brighter image (with amplified noise), and denoise the image using DDRM pretrained on ImageNet. 
We train an LDM~\cite{rombach2022high} from scratch using LOL and use 200 steps as prescribed.

\begin{figure*}[p]
\begin{center}
   \includegraphics[width=\linewidth]{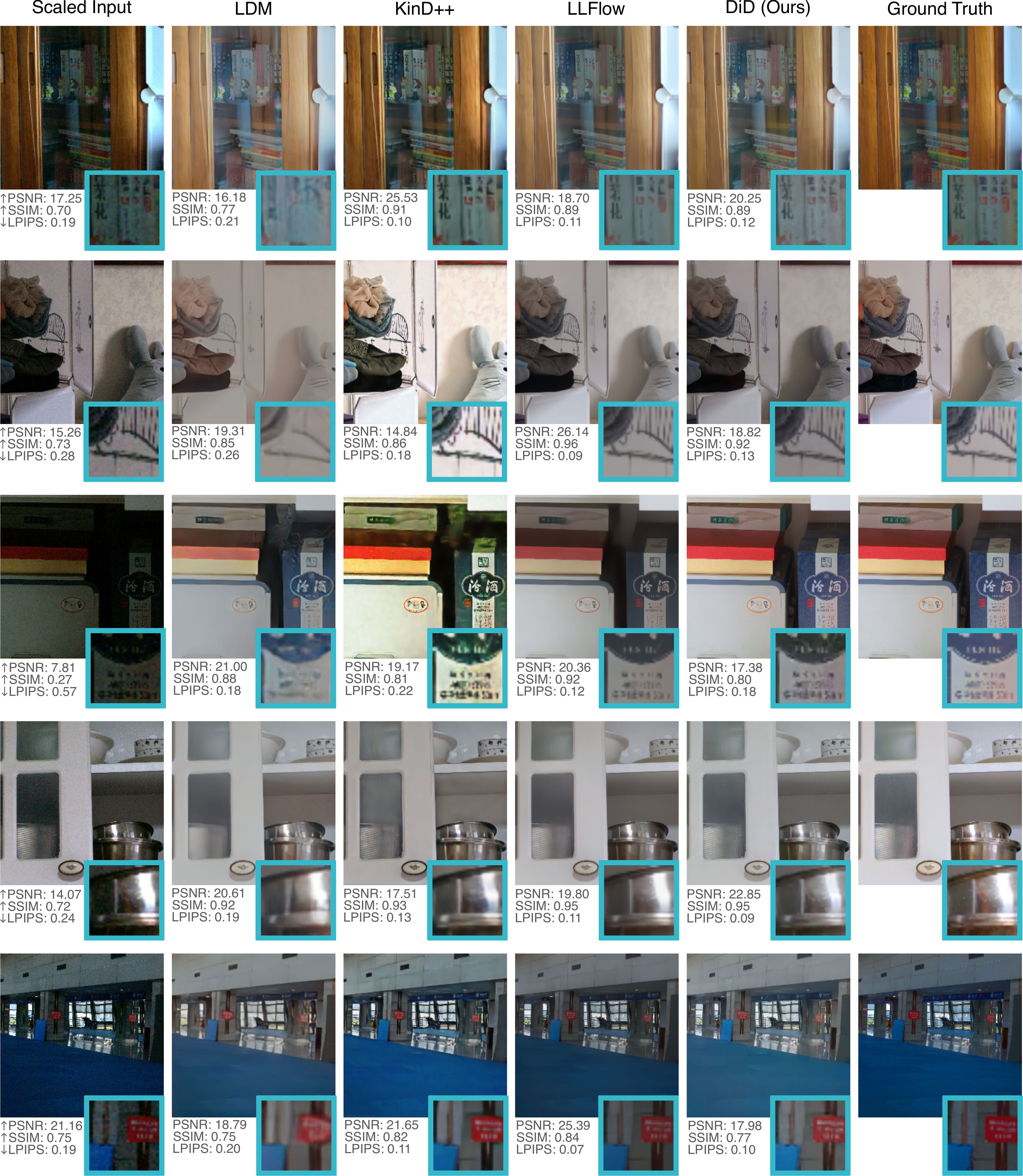}
\end{center}
\vspace{-0.6cm}
   \caption{\textbf{Qualitative results of baselines from more of the LOL test dataset.} We show results from top-performing low-light baselines. DiD reconstruction is competitive with reconstructions from other methods. We scale the input by a factor of 5 for visualization.}
\label{fig:squal1}
\vspace{-0.6cm}
\end{figure*}

We show more qualitative results from the LOL test dataset in Figure \ref{fig:squal1}.
DiD requires longer inference times than LLFlow on average due to the number of inference steps in the reverse diffusion process.
The same can be said about LDM. 
As faster sampling methods are being developed, as mentioned in our main paper, we believe the inference time for diffusion models can only be improved while maintaining better quality reconstructions than those from LLFlow.

\begin{figure*}[p]
\begin{center}
   \includegraphics[width=\linewidth]{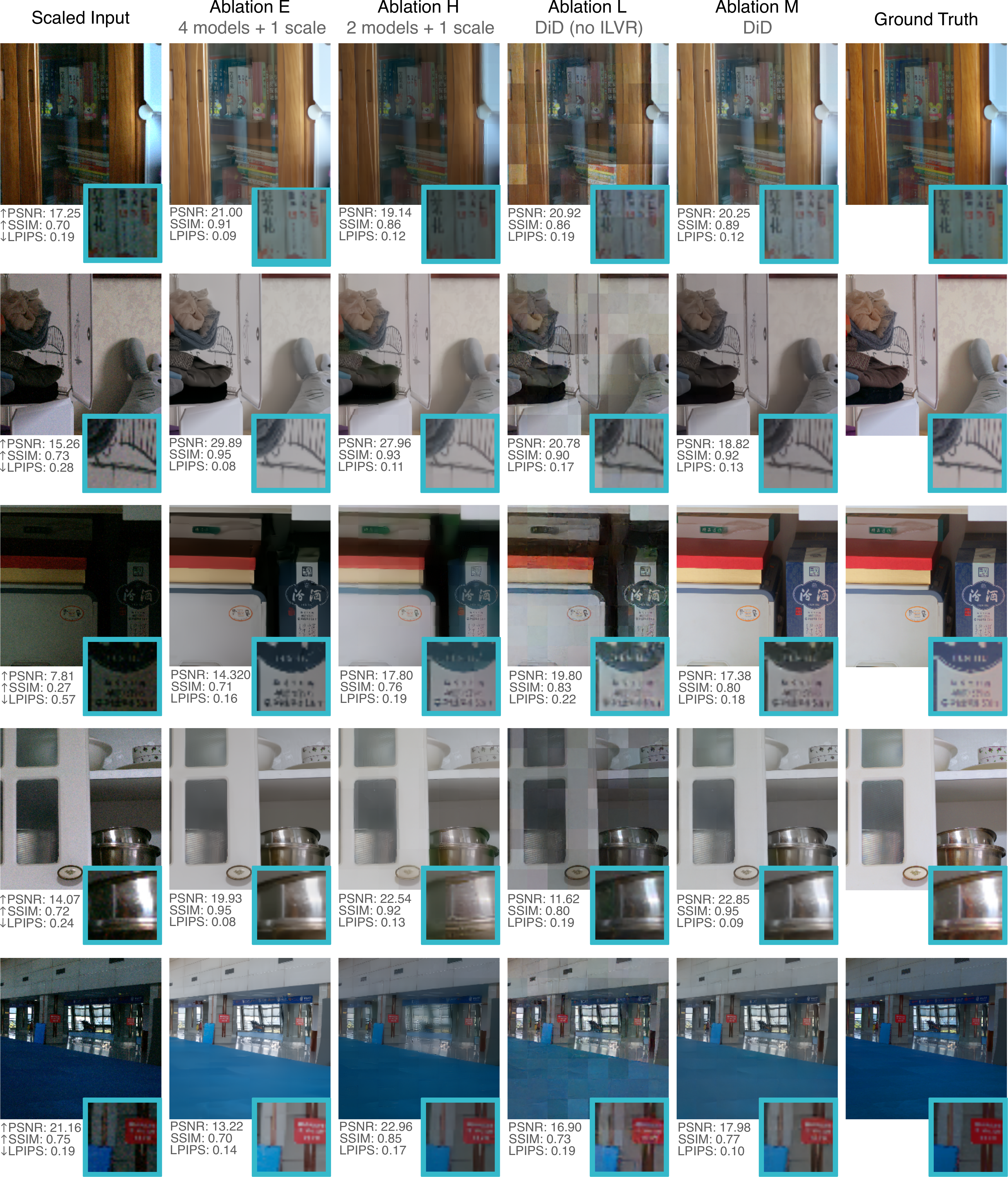}
\end{center}
\vspace{-0.6cm}
   \caption{\textbf{Qualitative results of ablations of the LOL test dataset.} We show results from top-performing ablations as described Table \ref{table:rest_ablations}. The combination of all described components, DiD performs the best robustly across images. We scale the input by a factor of 5 for visualization.}
\label{fig:squal2}
\vspace{-0.6cm}
\end{figure*}

\subsection{Ablation studies}
We clarify the term \textbf{model-to-scales} ratio.
A 4:1 model-to-scales ratio means that we trained 4 models. Each model is trained on 1 scale. An example of the 4 models using a 4:1 to ratio is as follows:
\begin{itemize}
    \item Model A is trained on $32\times32$ images that are downsampled versions of the $256\times256$ low-light measurement.
    \item Model B is trained on $32\times32$ patches that are taken from a $64\times64$ image (which is a downsampled version of the $256\times256$ low-light measurement).
    \item Model C is trained on $32\times32$ patches that are taken from a $128\times 128$ image (which is a downsampled version of the $256\times256$ low-light measurement).
    \item Model D is trained on $32\times32$ patches that are taken from the $256\times256$ low-light measurement.
\end{itemize}

\noindent A 2:1 ratio means we trained 2 models. Each model is trained on 1 scale. An example of the 2 models using a 2:1 to ratio is as follows:
\begin{itemize}
    \item Model A is trained on $32\times32$ images that are downsampled versions of the $256\times256$ low-light measurement.
    \item Model B is trained on $32\times32$ patches that are taken from the $256\times256$ low-light measurement.
\end{itemize}

\noindent A 1:2 ratio means we trained 1 model with 2 scales.
The model is trained on $32\times32$ patches that either are entire images from downsampling the low-light measurement from $256\times256$ to $32\times 32$ or are $32\times 32$ patches extracted from the original $256\times256$ low-light measurement.

We provide additional ablations highlighted in Table \ref{table:rest_ablations} and show qualitative results for top-performing ablations in Figure \ref{fig:squal2}.
All ablations use ILVR~\cite{choi2021ilvr} during inference unless specified otherwise.
For ablations, we report metrics computed on a randomly selected reconstruction rather than the best of 10 reconstructions.
We include an ablation study in which we attempt to refine the predictions in pixel space with a lightweight CNN.
This refinement network has 3 Conv2D+LeakyRELU layers with the following channel sizes $[3, 128, 3]$. 
We use an L2 loss and Adam optimizer for 10,000 iterations.
This CNN operates as a deterministic network to improve predictions. 
We train the CNN on $256 \times 256$ predictions from a pretrained DiD and compare the reconstruction to $256\times 256$ ground truth images.
However, we find that because the LOL test dataset has a significant distribution shift from its training dataset, there is an upper limit to how much the CNN can improve results. 
We observe comparable SSIM (\textcolor{green}{+0.06}), worse PSNR (\textcolor{red}{-0.47}), and comparable LPIPS (\textcolor{red}{+0.01}). 
Since the performance here was overall comparable, we instead report our original method DiD as part of our core contribution without any additional trainable parameters.

\setlength{\tabcolsep}{1.4pt}
\begin{table*}
\begin{center}
\caption{\textbf{Results from all ablation studies.} \textbf{Models/scales} refers to the number of trained models and the number of scales for which each model is trained. \textbf{Noise} refers to the addition of noise on the conditioning image. \textbf{LPIPS} refers to an additional LPIPS loss. \textbf{Data} refers to data normalization. \textbf{Cond.} refers to adding an upsampled scale 0 prediction to the conditioning input. Highlighted in blue are ablations which were not included in our main paper. We highlight the best and second best results using \textbf{bold} and \underline{underlined} text, respectively.}
\scalebox{0.93}{
\begin{tabular}{l|lcccc|ccc}
\hline\noalign{\smallskip}
ID & Models/scales & Noise & LPIPS & Data & Cond. & PSNR$\uparrow$ & SSIM$\uparrow$ & LPIPS$\downarrow$\\
\noalign{\smallskip}
\hline
\noalign{\smallskip}
A & 1 : 4 & \redmark & \redmark & \redmark & \redmark & 16.26 & 0.57 & 0.48\\
B & 1 : 4 & \redmark & \greenmark & \greenmark & \redmark & 19.56 & 0.74 & 0.35\\
C & 1 : 4 & \greenmark & \greenmark & \redmark & \redmark & 16.94 & 0.63 & 0.46\\
D & 1 : 4 & \greenmark & \greenmark & \greenmark & \redmark & 17.62 & 0.74 & 0.31\\
E & 4 : 1 & \greenmark & \redmark & \greenmark & \redmark & 19.63 & 0.80 & \textbf{0.14}\\
F & 1 : 2 & \greenmark & \greenmark & \greenmark & \redmark & 17.49 & 0.72 & 0.33\\
G & 1 : 2 & \greenmark & \greenmark & \greenmark & \greenmark & 18.37 & 0.73 & 0.33\\
H & 2 : 1 & \greenmark & \greenmark & \greenmark & \greenmark & 19.35 & 0.72 & 0.31\\

\rowcolor{LightCyan}
I & 1 : 4 & \greenmark & \redmark & \greenmark & \redmark & 17.78 & 0.74 & 0.31 \\
\rowcolor{LightCyan}
J & 2 : 1 & \greenmark & \greenmark & \greenmark & \redmark & 19.32 & 0.72 & 0.32 \\
\rowcolor{LightCyan}
K & DiD (with CNN) & \greenmark & \greenmark & \greenmark & \greenmark & \underline{20.53} & \textbf{0.88} & \underline{0.15} \\

L & DiD (no ILVR) & \greenmark & \greenmark & \greenmark & \greenmark & 17.78 & 0.72 & 0.36\\
M & DiD & \greenmark & \greenmark & \greenmark & \greenmark & \textbf{21.00} & \underline{0.82} & \textbf{0.14}\\
\hline
\end{tabular}

}
\label{table:rest_ablations}
\end{center}
\vspace{-0.5cm}
\end{table*}
\setlength{\tabcolsep}{1.4pt}

\section{Scene Text Recognition}
We simulate low light in scene text recognition by converting the images from RGB to HSV.
We then scale the V channel by a factor less than one (in our simulations, we use $0.4$ or $0.5$), following ~\cite{lv2021attention,zhang2021better} in simulating images under differing light conditions.
We follow the noise model from Mildenhall et al.~\cite{mildenhall2018burst}, which model Poisson-Gaussian noise as a Gaussian with zero-mean and signal-dependent variances.
We then convert the image back to RGB and add Poisson-Gaussian noise with a specified standard deviation for the Gaussian distribution and signal-dependent variance for the Poisson distribution.
We test the following datasets which display a wide range of capture quality:
\begin{itemize}
\item \textbf{IIIT5k-Words (IIIT5k) ~\cite{mishra2012scene}} which contains 3000 test images, most of which are of acceptable quality.
\item \textbf{ICDAR2013 (IC13)~\cite{karatzas2013icdar}} which consists of 1015 images for testing. The ICDAR 2013 and 2015 datasets are similar in text regularity and conditions.
\item \textbf{Street View Text (SVT)~\cite{wang2011end}} which consists of 647 images, many of which are severely degraded by blur, noise, and low resolution. 
\item \textbf{SVT-Perspective (SVTP)~\cite{phan2013recognizing}} which contains 645 images, with most suffering from heavy perspective distortion.
\end{itemize}

\begin{figure*}[t]
\begin{center}
   \includegraphics[scale=0.4]{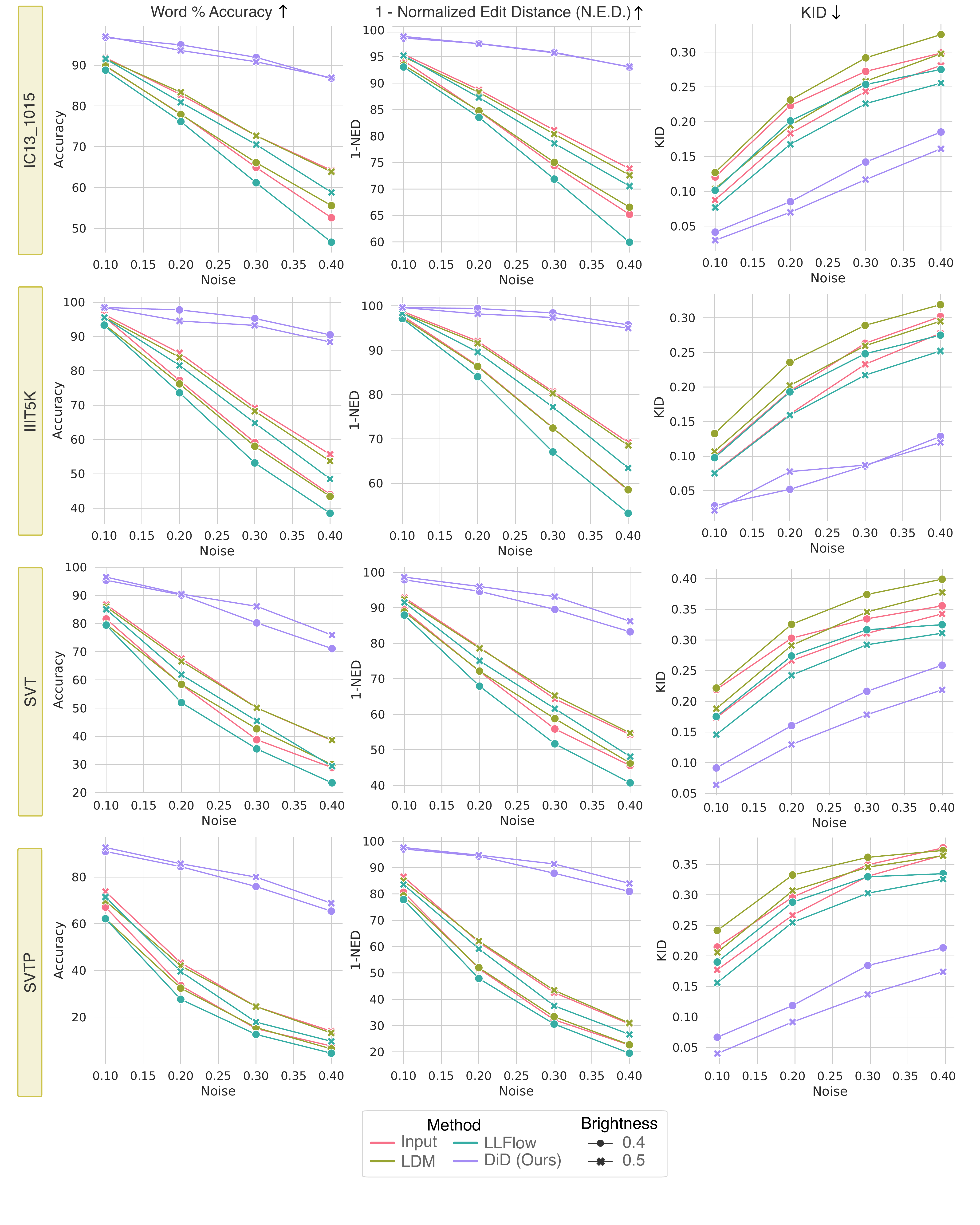}
\end{center}
\vspace{-0.6cm}
   \caption{\textbf{Quantitative performance on STR datasets.} We show performances of each method on each individual dataset at two levels of brightness and a range of Poisson-Gaussian noise levels using text recognition metrics (Word Accuracy and 1-Normalized Edit Distance) and KID. DiD performs robustly against noisy and dark conditions and exceeds in all these metrics.}
\label{fig:str1}
\vspace{-0.6cm}
\end{figure*}

\begin{figure*}[t]
\begin{center}
   \includegraphics[scale=0.45]{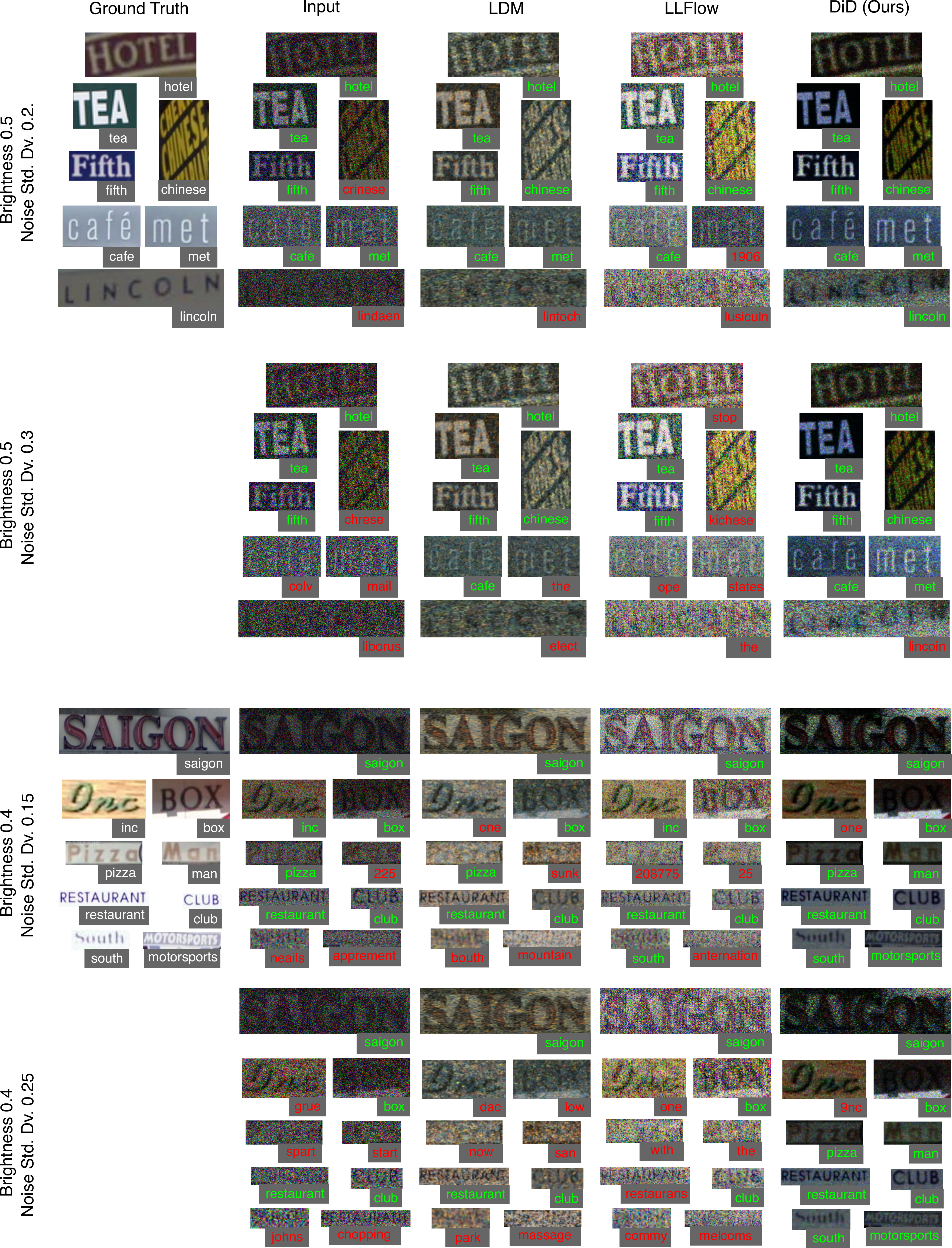}
\end{center}
\vspace{-0.6cm}
   \caption{\textbf{Qualitative results of STR performance on SVT dataset.} We show results of LDM~\cite{rombach2022high}, LLFlow~\cite{wang2022low}, and DiD on different examples in one of the four STR datasets. DiD is able to recover edges and high-frequency detail better in noisy and dark conditions to permit more accurate text recognition predictions than other methods can.}
\label{fig:str2}
\vspace{-0.6cm}
\end{figure*}

For each dataset, we sample 30 images to find the mean and standard deviation needed for tail-normalization.
For the text processing, we additionally scale our recovered image by 3. 
Since our method recovers an arbitrary exposure level without noise, scaling the image should not amplify any noise.
We show the performance of each method on individual datasets (a decomposition of Figure 5 from our main paper) in Figure \ref{fig:str1}.
We also show more qualitative results of different brightness and noise levels in Figure \ref{fig:str2}.

\section{Reconstruction on Other Datasets}

\begin{figure*}[t]
\begin{center}
   \includegraphics[width=\linewidth]{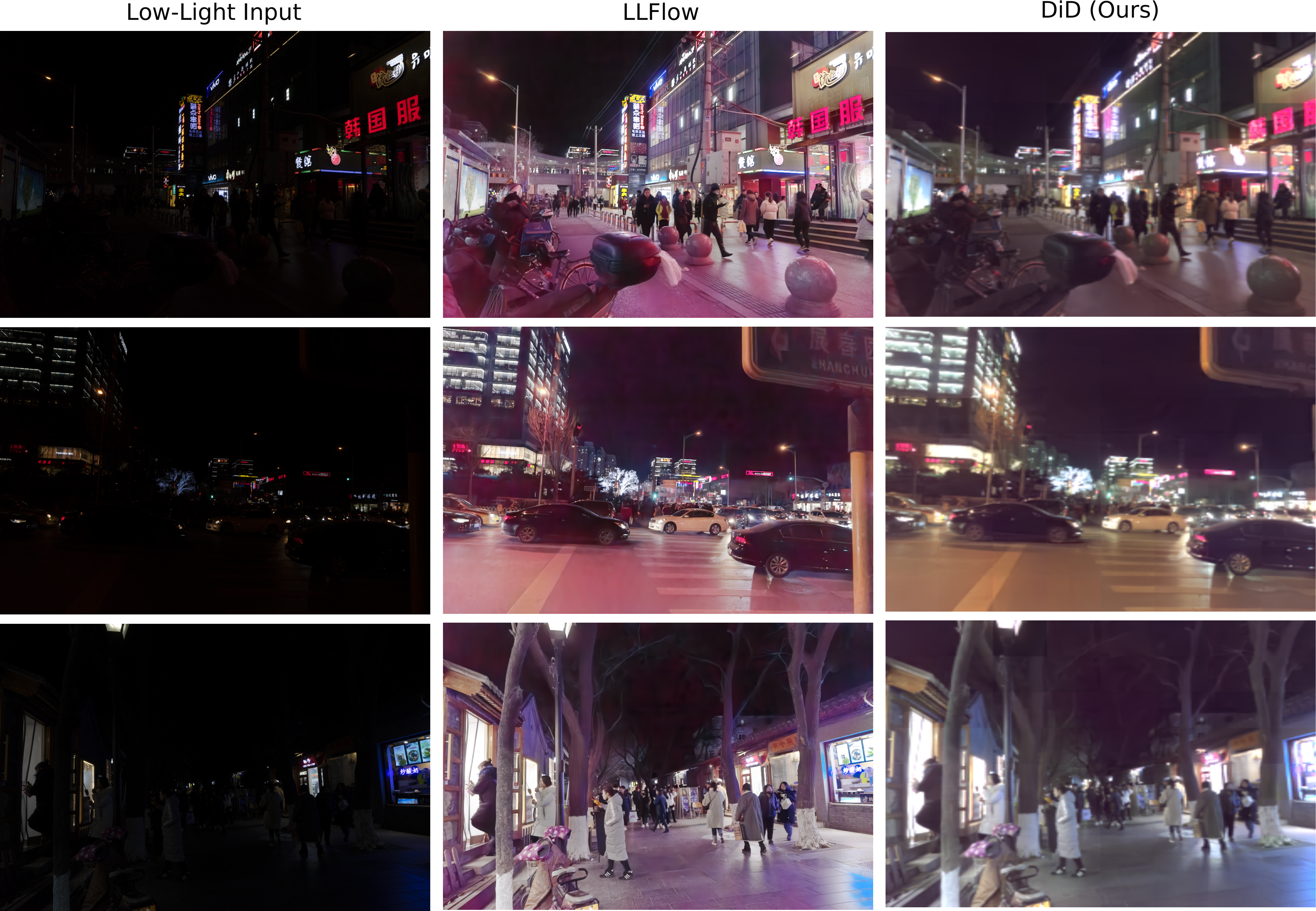}
\end{center}
\vspace{-0.6cm}
   \caption{\textbf{Reconstruction of DarkFace data, a real low-light task dataset.} DiD provides a realistic reconstruction of real low-light images, while LLFlow provides an unrealistic reddish tint. Both reconstructions could be used for face recognition, but DiD provides more aesthetically pleasing reconstructions.}
\label{fig:darkface}
\vspace{-0.6cm}
\end{figure*}

Many real low-light datasets are task datasets with no well-lit ground truth (Dark Zurich~\cite{sakaridis2020map}, ACDC~\cite{sakaridis2021acdc}, Nighttime Driving~\cite{dai2018dark}, CODaN~\cite{lengyel2021zeroshot}), so we cannot provide quantitative results on reconstruction performance.

Of the real low-light task datasets, only DarkFace~\cite{Chen2018Retinex} has been used for qualitative evaluation by 2 of 8 baselines (Zero-DCE and RUAS). 
We test our LOL-trained model on DarkFace (Fig. \ref{fig:darkface}), and found DiD to be highly robust against unseen, real test data, while LLFlow leaves an unrealistic red tint on images.

Our method could also be applied for other high-level downstream tasks such as segmentation and classification. 
However, our contributions are primarily in reconstructing high-frequency details, of which are not completely necessary for succeeding at segmentation and classification tasks.
We focus on instead on a task that requires high-frequency details, and thus shows the strengths of diffusion models.

\end{document}